\documentclass{aa}
\usepackage{graphicx, txfonts, natbib}
\usepackage[colorlinks=true,allcolors=blue]{hyperref}

\newcount\inclfigs \inclfigs=1

\def\kms{\hbox{km$\;$s$^{-1}$}} 
\def\arcsec{\hbox{$^{\prime\prime}$}}

\def\ha{\hbox{H${\alpha}$}}
\def\ca{\hbox{Ca~{\sc ii}}}
\def\aa{\hbox{AIA~193}}
\def\aac{\hbox{AIA~304}}

\begin{document}

\title{The chromospheric component  of coronal bright points }
\subtitle{Coronal and chromospheric responses to magnetic-flux emergence}

\author{Maria~S. Madjarska\inst{1, 2}, Jongchul Chae\inst{1}, Fernando Moreno-Insertis\inst{3, 4}, Zhenyong Hou\inst{5},  Daniel N\'{o}brega-Siverio\inst{6, 7}, Hannah Kwak\inst{1}, Klaus Galsgaard\inst{8}, 
Kyuhyoun Cho\inst{1}}

\offprints{madjarska@mps.mpg.de}
\institute{
Astronomy Program, Department of Physics and Astronomy, Seoul National University, Seoul 08826, Republic of Korea
\and
Max Planck Institute for Solar System Research, Justus-von-Liebig-Weg 3, 37077, G\"ottingen, Germany
\and
Instituto de Astrofisica de Canarias, Via Lactea, s/n, 38205, La Laguna, Tenerife, Spain
\and
Department of Astrophysics, Universidad de La Laguna, 38200, La Laguna, Tenerife, Spain
\and
Shandong Provincial Key Laboratory of Optical Astronomy and Solar-Terrestrial Environment, Institute of Space Sciences,
Shandong University, Weihai, 264209 Shandong, China
\and
Rosseland Centre for Solar Physics, University of Oslo, PO Box 1029, Blindern, 0315, Oslo, Norway
\and
Institute of Theoretical Astrophysics, University of Oslo, PO Box 1029, Blindern, 0315, Oslo, Norway
\and
School of Mathematics and Statistics, University of St Andrews, North Haugh, St Andrews, KY16 9SS, Scotland, UK
}

\date{Received date, accepted date}

\abstract
{We investigate the chromospheric counterpart of small-scale coronal  loops constituting a coronal bright point (CBP) and its response to a photospheric magnetic-flux increase accompanied by  co-temporal CBP heating.}
{The aim of this study is to  simultaneously investigate the chromospheric and  coronal  layers associated with a CBP, and  in so doing,  provide further understanding on the heating of plasmas confined in small-scale loops.}
{We used co-observations from the Atmospheric Imaging Assembly  and Helioseismic Magnetic Imager on board the Solar Dynamics Observatory, together with data from the  Fast Imaging Solar Spectrograph taken  in the \ha\  and \ca\ 8542.1~\AA\ lines. We also employed both linear force-free   and potential field extrapolation  models to investigate the magnetic topology of the CBP loops and the overlying corona, respectively. We used a new multi-layer spectral inversion technique to derive the temporal variations of the temperature  of the \ha\ loops (HLs).}
{We find that the counterpart of the CBP, as seen at chromospheric
  temperatures, is composed of a bundle of dark elongated features named in this work
  \ha\  loops, which constitute an integral part of the CBP loop
  magnetic structure. An increase in the photospheric magnetic flux due to
  flux emergence is accompanied by a rise of the coronal emission of the CBP
  loops, that is a heating episode. We also observe enhanced chromospheric
  activity associated with the occurrence of  new HLs and mottles. While the
  coronal emission and magnetic flux increases appear to be co-temporal, the response
  of the \ha\ counterpart of the CBP  occurs with a small delay of less than
  3~min.  A sharp temperature increase  is found in one of the HLs and  in
  one of the CBP footpoints estimated at 46\% and 55\% with respect to the
  pre-event values, also starting with a delay of  less than 3~min following
  the coronal heating episode. The  low-lying CBP loop structure
 remains non-potential for the entire observing period. The
magnetic topological  analysis {of the overlying corona} reveals the
presence of a coronal null point 
at the beginning and towards the end of the heating episode.} 
{The delay in the response of the chromospheric counterpart of the CBP suggests that the heating may have occurred at coronal heights.}

\authorrunning{Madjarska et al.}

\maketitle

\section{Introduction}
\label{intro}

Coronal bright points (CBPs) have been intensively studied for almost five
decades.  The CBPs are constituted by a set of small-scale coronal loops
with an average height of 6\,500~km that connect photospheric  magnetic-flux
concentrations of opposite polarity. As the plasma confined in these
loops is heated to a million degrees, they are seen with enhanced emission in
the extreme-ultraviolet (EUV) and X-ray wavelengths. The CBPs
 are found to be uniformly distributed in the quiet Sun
and coronal holes and  also appear in the vicinity of active regions.   For
a  review of CBPs, see  \citet{2019LRSP...16....2M}.

The CBP emission throughout EUV and X-ray wavelengths shows conspicuous
spatial and temporal variabilities \citep[for a review, see section~5.1
  in][]{2019LRSP...16....2M}. The first simultaneous observations of CBPs in
spectral lines formed at chromospheric, transition region, and coronal
temperatures were reported by \citet{1981SoPh...69...77H}. The study
revealed that the EUV emission from CBPs varies dynamically on
short timescales. The emission enhancements in coronal lines were
found to have a counter-response in chromospheric and transition-region lines. The
analysed spectroheliograms were taken with the Harvard EUV experiment on
Skylab/ATM at a 5\arcsec\ spatial resolution and 5.5~min cadence in the
C~{\sc ii}~1335~\AA, Ly-$\alpha$~1216~\AA, O~{\sc iv}~554~\AA, O~{\sc
  vi}~1032~\AA, C~{\sc iii}~977~\AA, and Mg~{\sc x}~625~\AA\ lines. From
observations taken near to the solar limb, it was found that the increase of the
coronal emission in CBPs precedes the lower temperature emission. The
authors therefore suggested that the heating possibly occurs at coronal heights and is
carried down to the chromosphere by thermal conduction. This is the first and
only study that has investigated the heating of CBP plasmas by exploring the
temporal response in spectral lines with various formation temperatures.
From the analysis of images taken with the SECCHI Extreme Ultraviolet Imager
(EUVI) on board the STEREO (Solar TErrestrial RElations Observatory) twin
spacecrafts in four passbands at 171~\AA, 195~\AA, 284~\AA, and 304~\AA,
\citet{2012ApJ...757..167K} concluded that CBPs are composed of hot loops
(logT (K) $\sim$ 6.2) overlying cooler loops (logT (K) $\sim$ 6.0) with cool
legs (logT (K) $\sim$ 4.9).

\begin{figure}[!ht]
\center
\ifnum \inclfigs >0
\hspace{-0.7cm}
\includegraphics[width=1.23\linewidth]{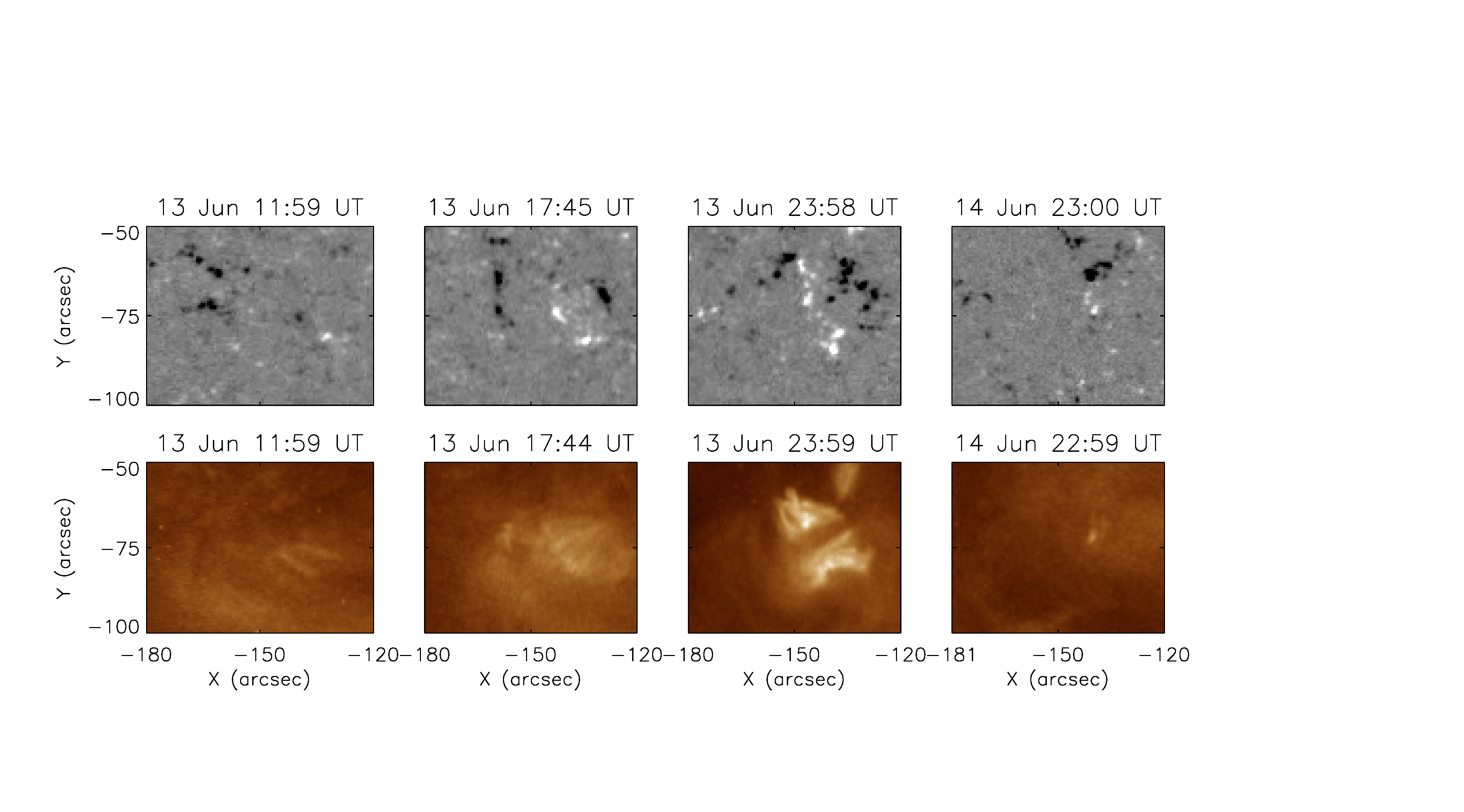}
\includegraphics[width=0.9\linewidth]{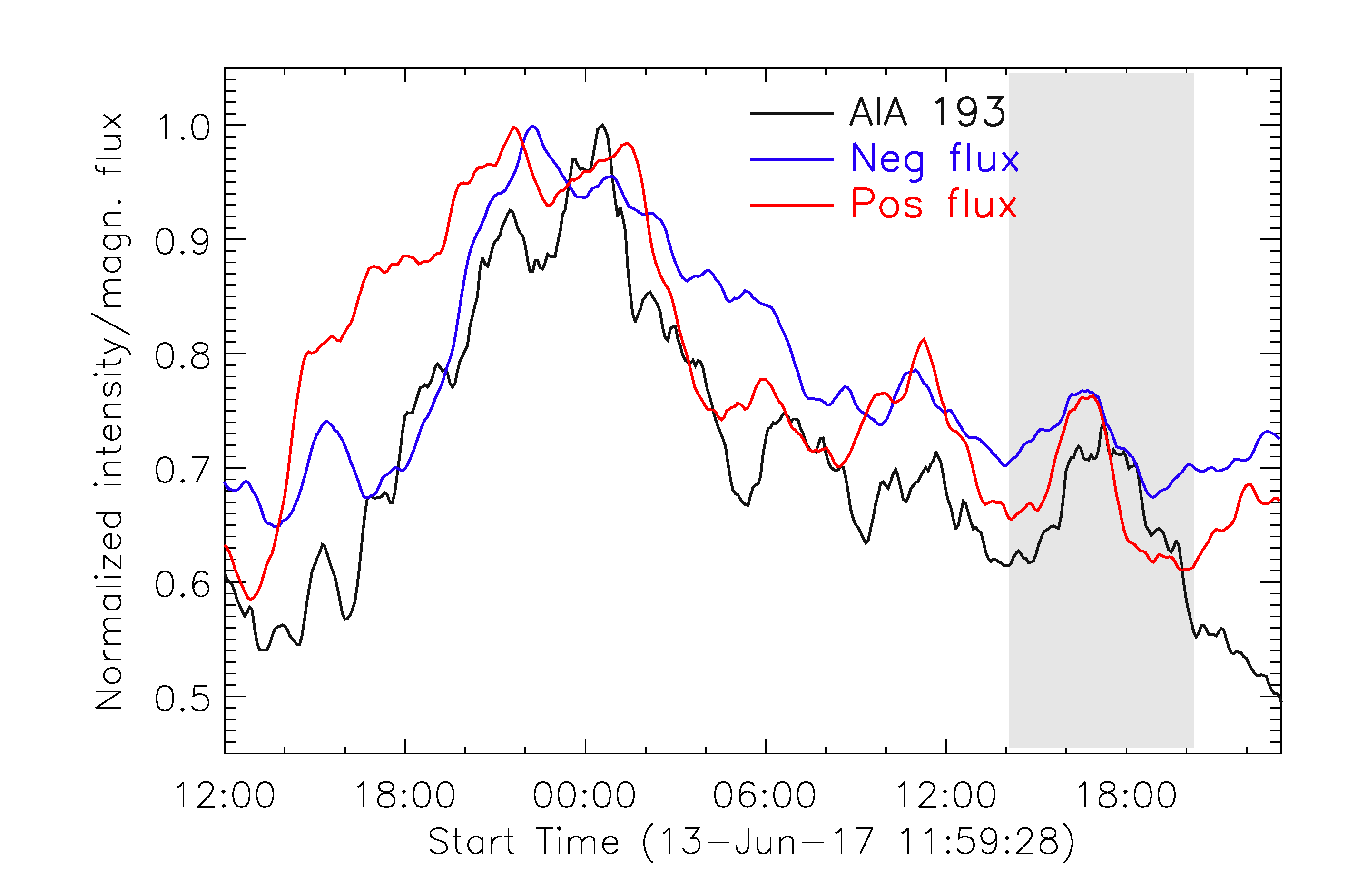}
\fi
\caption{Temporal variation of the magnetic flux and  \aa\ intensity of
    the CBP along its full lifetime. Images at the
      top: The HMI saturated at $\pm$100~G (top row) and \aa\ (bottom row).
Bottom panel: The \aa\ intensity and  LOS magnetic flux of the positive and negative
  polarities. Each curve is normalized to its
    maximum. The grey area
  highlights a period of magnetic flux  and coronal emission increase, part of
  which is covered by the FISS data and is analysed here in
  detail.}
\label{fig1}
\end{figure}

The Doppler velocities in CBPs were reviewed in detail by \citet{2019LRSP...16....2M}. In this work, we only recapture details that are relevant to the present study.  \citet{2003A&A...398..775M}  studied a single case of a CBP in a coronal hole using  raster-image observations in the S~{\sc vi}~933.40~\AA\  line (T $\approx$ 2.0 $\times$ 10$^5$~K maximum formation temperature)  taken with the Solar Ultraviolet  Measurements of  Emitted Radiation (SUMER) spectrometer on board the Solar and Heliospheric Observatory (SoHO). The  locations above the bipole magnetic-flux concentrations were associated with red-shifted emission (i.e. the footpoints of the CBP loops), while  blue-shifted emission was detected in the top or centre of the CBP loops. The blue-shifted signal was interpreted as coming from the underlying coronal hole rather than the CBP loops. We should not exclude the possibility that the emission from low-lying transition-region loops has also contributed to the blue-shifted emission. Because of  resolution limitations, the SUMER spectrometer could not resolve individual small-scale  loops, which is now possible  with the Interface Region Imaging Spectrograph (IRIS). 
\citet{2017SoPh..292..108K} analysed
two CBPs in data  obtained with IRIS as well as  with the Atmospheric Imaging Assembly (AIA)  and Helioseismic Magnetic Imager (HMI) on board the Solar Dynamics Observatory (SDO). The CBP loops were also seen  in the  IRIS slit-jaw images
(SJIs) taken in the Mg~{\sc ii}~k~2796 \AA\ (chromosphere) and C~{\sc ii}~1330 \AA\ (low transition
region) passbands.  Small-scale loops can be distinguished in the SJI C~{\sc ii} channel connecting bright-point-like sources that are the locations of the magnetic-flux concentrations. As in \citet{2003A&A...398..775M}, the study found  red-shifted emission in the transition-region  Si~{\sc iv} 1393.75~\AA\ (8.0 $\times$ 10$^4$~K) line  in the range 5--30~\kms, $\sim$5--20~\kms\ in  C~{\sc ii} 1394.76~\AA\ (2.5 $\sim$10$^4$~K), and $\sim$10~\kms\ in O~{\sc iv}~1401.16~\AA\ (1.6 $\times$ 10$^5$~K)  at the two
legs or footpoints of the CBP loops. Blue-shifted emission was recorded in the centre of the CBP. However,  the background emission from the coronal hole, as in the case studied by \citet{2003A&A...398..775M}, may also have contributed to this  emission. In coronal lines, \citet{2003A&A...399L...5X}   reported blue-shifted emission from a CBP in
the range 3--7~\kms\ in Ne~{\sc viii} 770.42~\AA\ (6.3 $\times$ 10$^5$~K)  without indication of the site of origin, while \citet{2008ApJ...681L.121T} obtained Doppler shifts close to zero in the same line. 

The chromospheric counterpart of CBPs has been poorly reported. An example of a study in the chromosphere of a CBP is the work  by  \citet{2009A&A...507.1625C}. An arch filament system (AFS) associated with a bipolar region seen in Ca~{\sc ii} 8542.1~\AA\ data taken with   the Interferometric Bidimensional Spectrometer (IBIS) on the Dunn Solar
Telescope (DST) of the United States National Solar Observatory at Sac
Peak was analysed. The AFS was found to connect a bipolar region and to have a
counterpart at coronal temperatures  in images taken with the Extreme ultraviolet Imaging
Telescope (EIT) on board SoHO. In view of the statements by the authors,
we can conclude that this region represents a CBP. The CBP  lifetime was $\sim$3.5 days. The CBP  formed following a bipole flux emergence.  The magnetic fragments  evolved through various processes including divergence, convergence, and cancellation in the latest stage of the bipole lifetime.  The velocities were  derived from the shift of the Ca~{\sc ii} line centroid  and cloud model flows (see the paper for details on the methodology). While the velocities obtained from either method are similar at the loop top, these velocities are different  in the loop footpoints. The study reports that  the line-of-sight (LOS) velocities are similar with values that are greater by a factor 3--4 for the Doppler shift method.
In summary, red-shifted emission is known to dominate at the footpoints or legs of CBP loops. Blue-shifted emission is observed at  the centre or top of the loops at transition-region temperatures, but the contribution from the
  background emission of the coronal hole can be 
  important, especially in spectral lines formed at higher transition region and coronal temperatures.

Magnetic flux emergence from the convection zone into the overlying
  solar atmosphere is an essential physical process  that has a fundamental impact in the activity of the
    solar atmosphere on a wide range of length scales \citep[for a review
    see][]{2014LRSP...11....3C}. Magnetic-flux emergence in the quiet Sun
  causes the formation of CBPs in around 50\% of all observed cases
  \citep[e.g.][and references therein]{2018A&A...619A..55M}. Typically 30
  min to 60 min after a visual detection of the emerging bipolar flux in the
  photosphere, the first CBP coronal loops can be detected in \aa, often
with projected size as small as 5\arcsec.  In active regions
  magnetic flux emergence is linked to the formation of various eruptive
  phenomena, including Ellermann bombs \citep[e.g.][and references
    therein]{2019A&A...626A..33H}, 
surges \citep[e.g.][and references therein]{2017ApJ...850..153N},
filament 
  eruptions, and coronal mass ejections \citep[e.g.][and references
    therein]{2018ApJ...862..117D}.

In the present paper we investigate the temporal evolution of a CBP as seen
  in the \aa~\AA\ channel of SDO in parallel with the simultaneous evolution
  in morphology, dynamics, and temperature obtained from chromospheric lines  (\ha\ and \ca\ 8542.1~\AA) 
 based on data taken with the Fast Imaging Solar Spectrograph (FISS).  A CBP brightness increase is observed in the 193~\AA\ channel of SDO/AIA  and is associated with a photospheric magnetic-flux
  increase recorded in  LOS SDO/HMI magnetograms.  For the
  first time, magnetic field lines obtained from  a  linear force-free field  (LFFF)  extrapolation model
  are matched with CBP loops seen in images taken with the AIA~193~\AA\ filter  and their parameters are derived.  The paper is organized as follows. Section~\ref{obs_met} reports on the analysed observational material and the methodology used to derive various physical parameters.  The results are given in Section~\ref{res.obs}. The discussion is presented in Section~\ref{disc} and the summary and  conclusions are in Section~\ref{concl}.

\section{Observational material and methodology}
\label{obs_met}

The chromospheric observations were taken with FISS at the Big Bear Solar Observatory on 2017 June 14.  The  FISS is
a dual-band \'{e}chelle spectrograph recording simultaneously the \ha\ and
the \ca\ 8542.1~\AA\  lines \citep{2013SoPh..288....1C}. The images are produced in a raster mode with a 0.16\arcsec\ wide slit. The data are obtained using an adaptive optics system. The spectral sampling is 0.019~\AA\ for the \ha\ line and 0.026~\AA\ for the \ca\  line. The raster cadence is $\sim$27~s with a raster time step of 0.13~s. The data were taken in the quiet Sun. The images in the far wings of   \ha\ and \ca\ were used to align with the HMI magnetograms. The field of view (FOV) is 24\arcsec\ $\times$ 40\arcsec\ (1.74 Mm $\times$ 29 Mm) and it is tilted with respect to the north-south direction.  Only observations taken between 17:52:03~UT and 18:49:33~UT were used for the analysis as they had the best quality.  

To obtain the Doppler velocities we used the  lambdameter method that is illustrated and described in \citet[][and references therein]{2014ApJ...789..108C}. The obtained Doppler velocities are not absolute values as they are obtained with respect to a reference spectrum from a region away  from the CBP (see section~\ref{res.obs} for more details). Nevertheless the Doppler shifts  give a correct image of the velocity pattern and its evolution in time.  The hydrogen temperature at the chromospheric heights under the CBP is obtained from a new multi-layer spectral inversion technique developed by \citet{2020A&A...640A..45C}.

To investigate the coronal part of the CBP, we analysed imaging data taken by
AIA \citep{2012SoPh..275...17L} on board SDO \citep{2012SoPh..275....3P}. We
used AIA data taken in the 193~\AA\ (hereafter \aa) and 
  304~\AA\ (hereafter \aac) channels. The \aa\ channel is dominated by
  Fe~{\sc xii} lines (log T(K) $\sim$ 6.2). The \aac\ channel is dominated by
  the two He~{\sc ii} lines at 303.786 and 303.781~\AA\ (log T(K) $\sim$ 4.7),
  but this channel also has a contribution from the coronal Si~{\sc xi} at
  303.33~\AA\ (log T(K) $\sim$ 6.2) \citep{2010A&A...521A..21O}. The
  formation of the He~{\sc II} lines is complex \citep{2012SoPh..279...53A,
    2017A&A...597A.102G} and the \aac\ channel  is known to
  predominantly show the footpoints of coronal loops as well as in CBP loops. The
  contribution by Si~{\sc xi} at 303.33~\AA\ is estimated to be at the 11\% level in
  active regions  \citep{2010A&A...521A..21O} implying  that the
  \aac\ channel may also show a  contribution from the coronal emission in CBPs.  Chromospheric low-temperature features appear dark in these channels (but also in other coronal
channels). This is caused by the extinction of photons emitted from the hot
coronal plasma in overlying neutral H~{\sc i} or He {\sc i} atoms in 
cool plasma \citep{1999ASPC..184..181R}. The AIA EUV data have a 12~s cadence and
0.6\arcsec\ $\times$ 0.6\arcsec\ pixel size.

The HMI LOS magnetograms  \citep{2012SoPh..275..207S}  used in this study
have either  a 45~s or a 720~s cadence and 0.5\arcsec\ $\times$ 0.5\arcsec\ pixel size. The magnetograms with 720~s cadence were used for the LFFF extrapolation because of their higher signal-to-noise ratio. The 45~s cadence data were used for the co-temporal FISS study. The magnetic field data were corrected for the projection effect. The AIA 1600~\AA\ was used  to align the HMI data with those from \aa~\AA.   The precision of the alignment between the different instruments is $\sim$1\arcsec. To  produce the curves of the positive and negative fluxes over the CBP full lifetime we used  a 5~min data-cadence sequence. All images in this work were rescaled to the FISS pixel size, which is 0.16\arcsec. The \aa, \ha, and \ca\ curves were smoothed using a seven-point window, while for the HMI data,  a four-point window was applied.


\begin{figure*}[!ht]
\centering
\includegraphics[scale=1.]{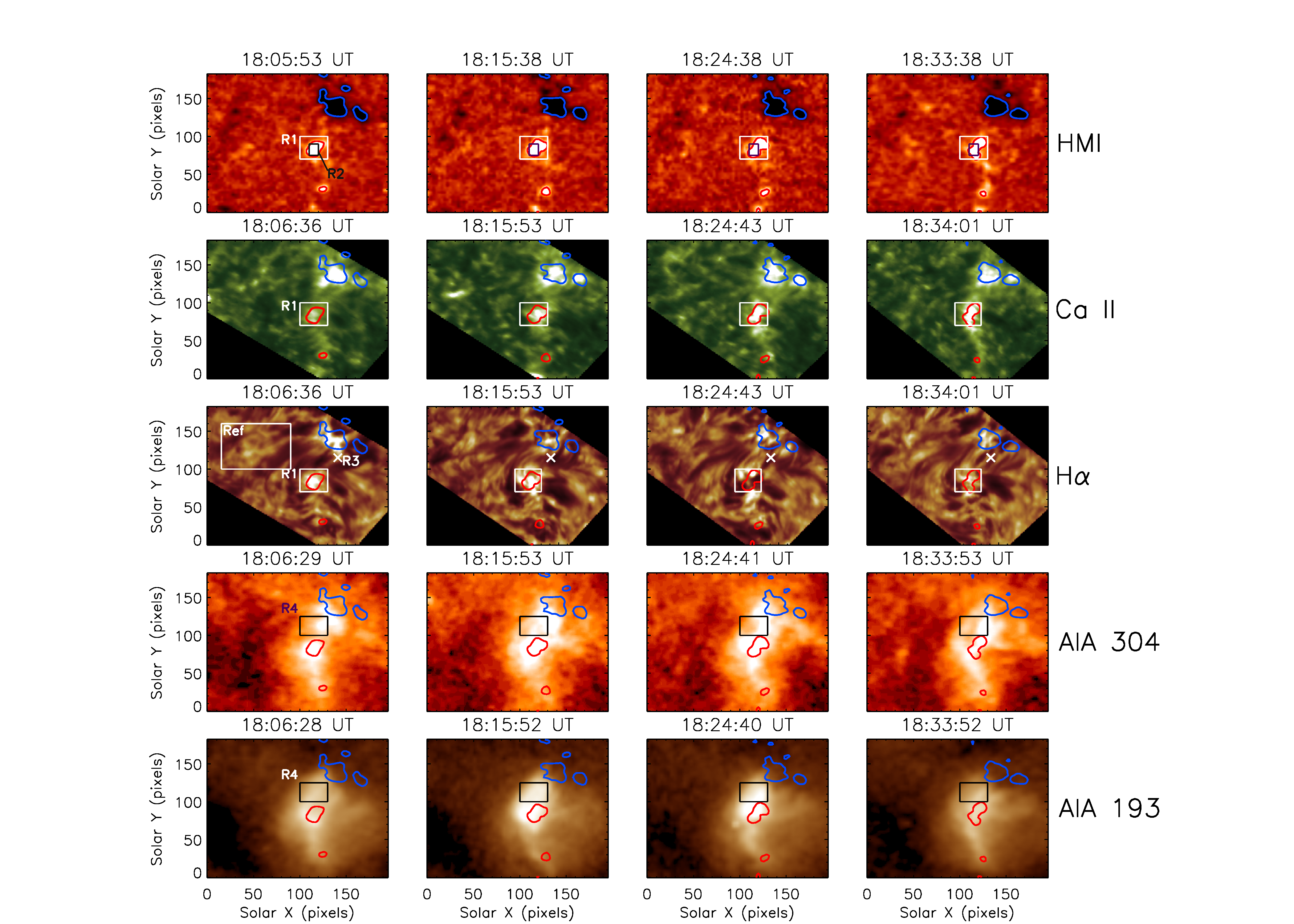}
\caption{Multi-instrument images of the CBP. Top row:  HMI line of sight magnetograms scaled from $-$50 to 50~G  overplotted  with a larger square noted as region of interest \#1 (R1, also on the \ha\ and \ca\ images), which outlines the area from which the temporal variations of various parameters in the south footpoint of the CBP were obtained. The smaller square denotes
  (R2) the area from which the hydrogen temperature was determined in
  Fig.~\ref{fig7}. Second row: The \ca\ line-core intensity  images
  overplotted with the $\pm$50~G magnetic-field contours. Third row: The
  \ha\  line-core intensity images with the same contours. The   large square on the
  \ha\  image in the first column (Ref) is the area from which  the
  reference \ha\ profile used to derive the Doppler velocities was  obtained.
  The cross sign denotes the centre of a small area (R3) from which the
  temporal variations in an \ha\ loop were produced in Figs.~\ref{fig5} and
  Fig.~\ref{fig7}. Fourth and fifth rows:
 The \aac\ and \aa\ images overplotted with the $\pm$50~G magnetic-field contours and a square (R4) that outlines the area from which the light curve in the CBP loops was produced for Figs.~\ref{fig5}, \ref{fig6}, and \ref{fig7}. The pixel size is 0.16\arcsec, that is the size of the images is $29\arcsec  \times 31\arcsec$.}
\label{fig2}
\end{figure*}

We investigated the topology of the magnetic field, both of the
low-lying loops and of the overlying coronal structure. To
obtain the magnetic structure of the CBP loops, we first processed the
\aa\ images using the multi-scale Gaussian normalization (MGN) code
developed by \citet{2014SoPh..289.2945M}.  Next, we applied a LFFF
extrapolation model using the  LOS magnetic-field measurements from
HMI.  A low average plasma-$\beta$ allowed us to assume that to the lowest
order the magnetic field is force-free; that is the current density is
aligned with the magnetic field. Details concerning the methodology can be found in
\citet{2002SoPh..208..233W} and \citet{2010ApJ...723L.185W}. The LFFF
contains the force-free parameter $\alpha$ (m$^{-1}$). In this work, the dimensionless quantity $\alpha$L is used, 
where $L$ is the harmonic mean of  the sides, $L_x$ and $L_y$, of the magnetogram (i.e. $1/L^2 =
1/2(1/{L_x}^2 + 1/{L_y}^2))$.  We carried our computations with  $\alpha$L values in the range from $-$4.0 to 4.0, 
with a step of 0.5. The MGN
image processing helps to enhance the CBP loops so that a visual selection
could be made along their entire length. Then, a fit was
attempted between the LFFF extrapolation and the visual identification from
the MGN image processing. The least-squares method was used to find the
field line, which has the minimum distance between the interpolated field
lines and the loop visually identified by us.

Additionally, a potential field extrapolation model was applied to the
magnetograms taken at 17:21 UT, 18:10~UT, 18:15~UT, and 18:25~UT, to
investigate the magnetic topology of the overlying corona and to search for
the presence of magnetic null points. Details of the extrapolation
methodology are given in section~2.2 of \citet{2017A&A...606A..46G}, while the
methodology of the magnetic topological analysis is described in section~2.3
of the same paper.

\section{Results}
\label{res.obs}

\subsection{CBP formation and evolution}

In the search for suitable observations we were very fortunate to discover
that the highest quality FISS data in the quiet Sun are co-spatial and
  co-temporal with a CBP despite the FISS small FOV of 24\arcsec\ $\times$
40\arcsec. The initial data analysis was even more
encouraging as it showed that during the one hour of the highest-quality
observations, the CBP had an episode of intensity increase recorded in the
 \aac\ and 193 channels that was linked to a rise of the photospheric magnetic
flux in the footpoints of the CBP loops.

The CBP was formed from bipolar flux emergence, which is typical for at least 50\% of the CBPs \citep{2018A&A...619A..55M}. The emergence started at $\sim$13:35~UT on 2017  June 13 (Fig.~\ref{fig1}). The first signature of the CBP in \aa\ is at 14:44~UT on June 13 and the last time a trace of the CBP is visually detectable in this channel is on June 14 at $\sim$23:00~UT, which gives  a lifetime of  $\sim$33~hrs.  The CBP positive polarity fully vanishes at 02:00~UT on June 15, while a patch of the negative flux remains.  Figure~\ref{fig1} (top row images) and Figure~\ref{fig2} (top row images)  contain the HMI magnetograms, which clearly show the opposite polarities
  where some of the CBP loops are rooted (see section~\ref{topology} for details). The general evolution of the photospheric  magnetic flux follows a typical evolutionary pattern seen in CBPs forming from flux emergence \citep[e.g.][]{2018A&A...619A..55M}.  The flux emergence proceeds with a continuously diverging bipole until the two polarities reach a certain distance, while the CBP attains its largest size and brightness.  Thereafter, a magnetic flux convergence process follows until the full disappearance of one of the polarities (the positive one).  Occasionally flux coalescence and cancellation are observed. The cancellation is most prominent towards the end of the lifetime of the CBP.  During the magnetic flux emergence the positive flux approaches a pre-existing negative flux, which leads to  micro-flaring and to a mini-ejection (this event will be the subject of investigation in a follow-up study). During the specific period of the FISS observations, a displacement of $\sim$0.27$\pm$0.03~\kms\ of the positive flux was approximately estimated while the negative flux remained almost at the same location.

\begin{figure*}[!ht]
\center
\ifnum\inclfigs>0
\includegraphics[scale=0.6]{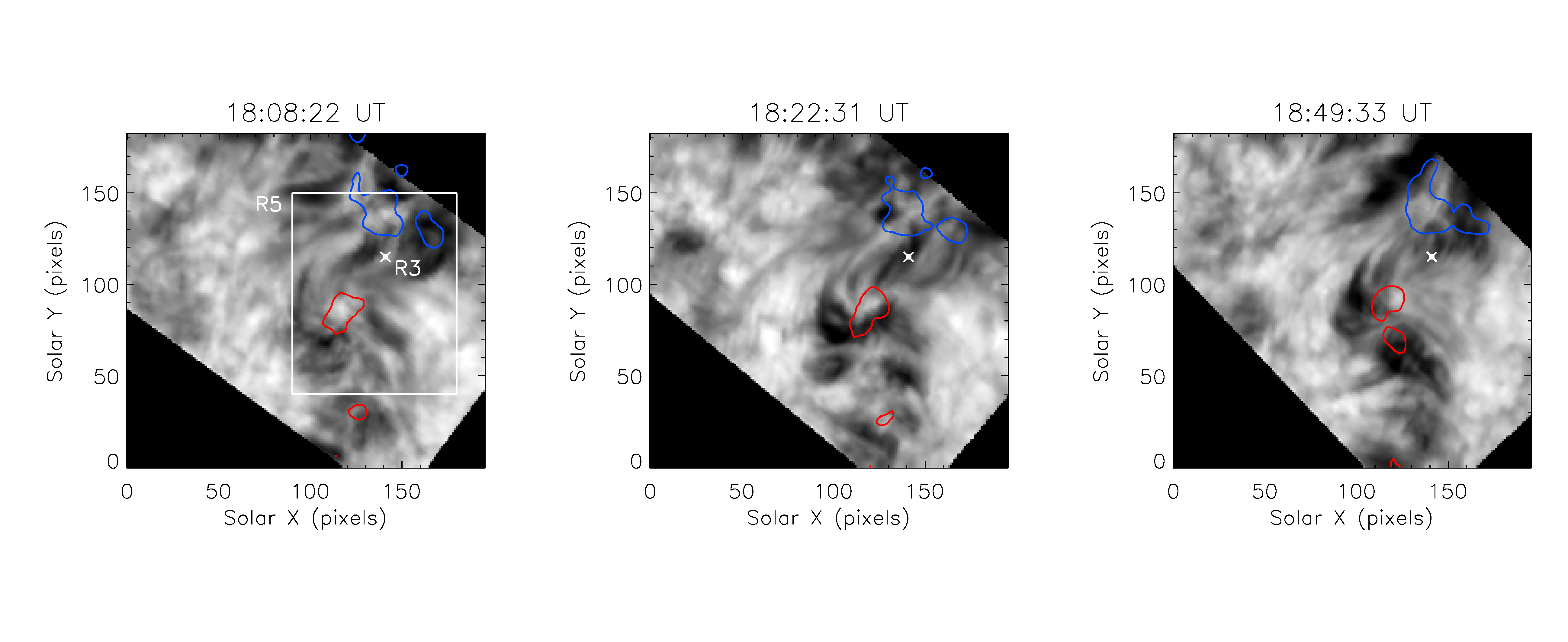}
\fi
\caption{\ha\ images before (left), during (middle) and after (right) the heating episode with overplotted the $\pm$50~G magnetic-field contours (red -- positive polarity, blue -- negative). The images are produced by summing the \ha\ intensity in the blue and red wings of \ha\ at $\pm$0.45~\AA.  The large square  denoted as R5 in the left panel is where the temporal variations from the  whole CBP shown in Figs.~\ref{fig5} and \ref{fig7} were obtained. A cross denotes R3, from which the temporal variations in one of the {\ bf HLs} were produced in Figs.~\ref{fig5} and \ref{fig7}. The pixel size is 0.16\arcsec.}
\label{fig3}
\end{figure*}
  We paid attention to the full lifetime evolution of the coronal intensity,
  as recorded in \aa,\ and to the bipolar magnetic flux.  As in all emergence
  cases shown in \citet{2018A&A...619A..55M}, there is a strong correlation
  between the temporal evolution of the coronal intensity recorded in
  \aa\ and the photospheric magnetic flux.   In Fig.~\ref{fig1} we show the temporal variations of the normalized \aa\ intensity
  of the CBP over its lifetime, and the normalized total positive and
  negative magnetic fluxes.  In the top panel of
    Fig.~\ref{fig1} we show two rows (top -- HMI, bottom -- \aa)
    consisting of four images each that
    display  the time evolution of the CBP, including its appearance, the
      phase of maximum size and intensity, and its disappearance. The window is slightly offset from
    the centre of the CBP to catch  the CBP loops that extend
      to the east, since the CBP is not observed in the disc centre.  The curves in the bottom panel of     Fig.~\ref{fig1}
  were produced using the 5~min cadence data and from a FOV slightly larger
  than the one shown in Fig.~\ref{fig2} in order to enclose the CBP while
  reaching its maximum size earlier in its lifetime ($\sim$00:00~UT on June
  13). A heating event occurred between $\sim$18:07~UT and $\sim$18:33~UT on
  June 14, that is towards the end of the CBP lifetime, which we analyse in
  the following sections. The curves in Fig.~\ref{fig1} reveal that the
  period of time covered by the FISS data is part of a longer lasting heating
  event following a magnetic-flux increase that started around 14:00~UT on
  June 14 (shown with a grey background).  As is generally the
    case for the whole evolution of the CBP,  there is a strong correlation between the magnetic flux increase 
  (emergence) and enhancement of the coronal emission. The heating event
  lasted a bit less than 3~h. For the full lifetime of the CBP, the linear
  Pearson correlation coefficient for the positive flux and the
  \aa\ intensity is 0.75, while for the negative flux it is 0.88. As not all
  the flux in the FOV is involved in the CBP magnetic configuration, the
  correlation coefficient is at least 0.9.

\subsection{Chromospheric and coronal morphology of the CPB}
\label{morph}
The \ha\ line-core intensity  images in  (the third row of
Fig.~\ref{fig2}) show the CBP
chromospheric morphology, which can simply be described as a bundle of
slightly curved loops (only the southern end  is sigmoid shaped) that connect
a bipolar region. The \ha\ line-core intensity was derived from the
lambdameter method  (see section~\ref{obs_met}).  The loops are seen as
elongated dark features that lie along the path between the magnetic
polarities of opposite flux. In  Fig.~\ref{fig3} we show images that were
produced by summing the intensity flux in the wings of \ha\ at
$\pm$0.45~\AA,\ where  the structure of the  evolving loops of the CBP
can best be followed.   Hereafter we use the term \ha\ loops (HLs)
  of the CBP, as these features are certainly part of the whole CBP
magnetic-plasma complex. Often studies refer to these phenomena
in the quiet Sun as arch
filaments \citep[e.g.][and references therein]{2009A&A...507.1625C} or
fibrils \citep[e.g.][and references therein]{2012ApJ...749..136L}, while
in active regions they are named fibrils
\citep[e.g.][]{2008A&A...480..515C,2012SSRv..169..181T}.   

The second and third rows of panels in Fig.~\ref{fig2}
allow us to compare the line-core intensity of the \ca\ line (second row)
  with those of \ha\ just mentioned. The footpoints of the HLs are rooted in the magnetic-flux
  concentrations of opposite polarity and appear as bright features in both
  the \ha\ and the \ca\ images as typically seen above magnetic-field
  concentrations. In the line wings the brightenings have sharp edges while
  in the line centres they appear larger and more diffuse at the edges
  because of the lateral divergence of the magnetic field with height.  In
  addition to the HLs, we observe dynamically evolving absorption features
  that both rise and fall back on short timescales (a few minutes). These features are
  known as mottles \citep[for details on mottles see][and references
    therein]{2012SSRv..169..181T} and are best seen in the wings of the
  \ha\ line (see the animation in Fig.~\ref{figa1}) emanating from the
  magnetic concentrations, which also host the footpoints of the CBP loops
  \citep[e.g.][]{2009A&A...507.1625C}. Mottles are known to propagate both
  along closed and open magnetic field lines \citep{2012SSRv..169..181T}. We
  also observe up- and downflows in the legs of the HLs. These flows possibly
  would have been identified as mottles by a mottle-dedicated study, but
  hereafter we consider them as a separate feature because of their
  association with the CBP magnetic structure.  A closer look at the
  footpoints of the CBP loops in the \aa\ images also reveals these dark
  (absorption) dynamically evolving features (see the animation in
  Fig.~\ref{figa1}). The combination of \ha\ and \aa\ images clearly
  demonstrates that mottles and cool plasma upflows in the legs of the CBP
  loops are responsible for the darkening in the legs or footpoints and the
  adjacent areas of the CBP loops.  This causes for instance the absence of
  coronal emission above the northern CBP footpoint at the location of the
  negative polarity creating the illusion of the coronal emission from the
  CBP to be misaligned with respect to the magnetic polarities.  In the south
  footpoint (positive magnetic polarity) the loops are sigmoidal and
  therefore do not create such an obscuration.  Often the
  mottle or cool-plasma-flow evolution also creates small-scale temporal and
  spatial variations of the coronal intensity. In reality, these are the
  absorption features that evolve dynamically along the CBP loop legs and
  even rise up along the loops. This can be followed in the provided
  animation in Fig.~\ref{figa1}. { In the \aac\ images in Fig.~\ref{fig2}
    we can clearly distinguish the bright footpoints of the CBP as well as
     the so-called legs (the low part of the CBP loops), which is common when CBPs
    are observed at transition-region temperatures \citep[for details
      see][]{2019LRSP...16....2M}.  The CBP loops also become visible in this
    channel during the analysed observing period. We come back to this later
    in the next section. The north CBP footpoint is also obscured by mottles
    as in the coronal \aa\ channel. Chromospheric features such as filaments and
    mottles or spicules are usually seen in absorption when observed on the disc
    in \aac\ \citep[e.g.][and references therein]{2020A&A...643A..19M}.}

While the HLs evolve more dynamically, the coronal loops show little change (see the animation in  Fig.~\ref{figa1}). This, however, may be due to the different spatial resolution of the two instruments, which is 0.16\arcsec/px for the FISS chromospheric data and 0.6\arcsec/px for the AIA data. The cross-section of several HLs  were measured and an average width of  $\sim$0.41\arcsec, with a minimum size of 0.28\arcsec\ and maximum of 0.56\arcsec\ , were obtained. These were determined from the FWHM of their absorption profile cross-section. 

\begin{figure*}[!ht]
\center
\vspace{-2.5cm}
\ifnum\inclfigs>0
\includegraphics[scale=0.8]{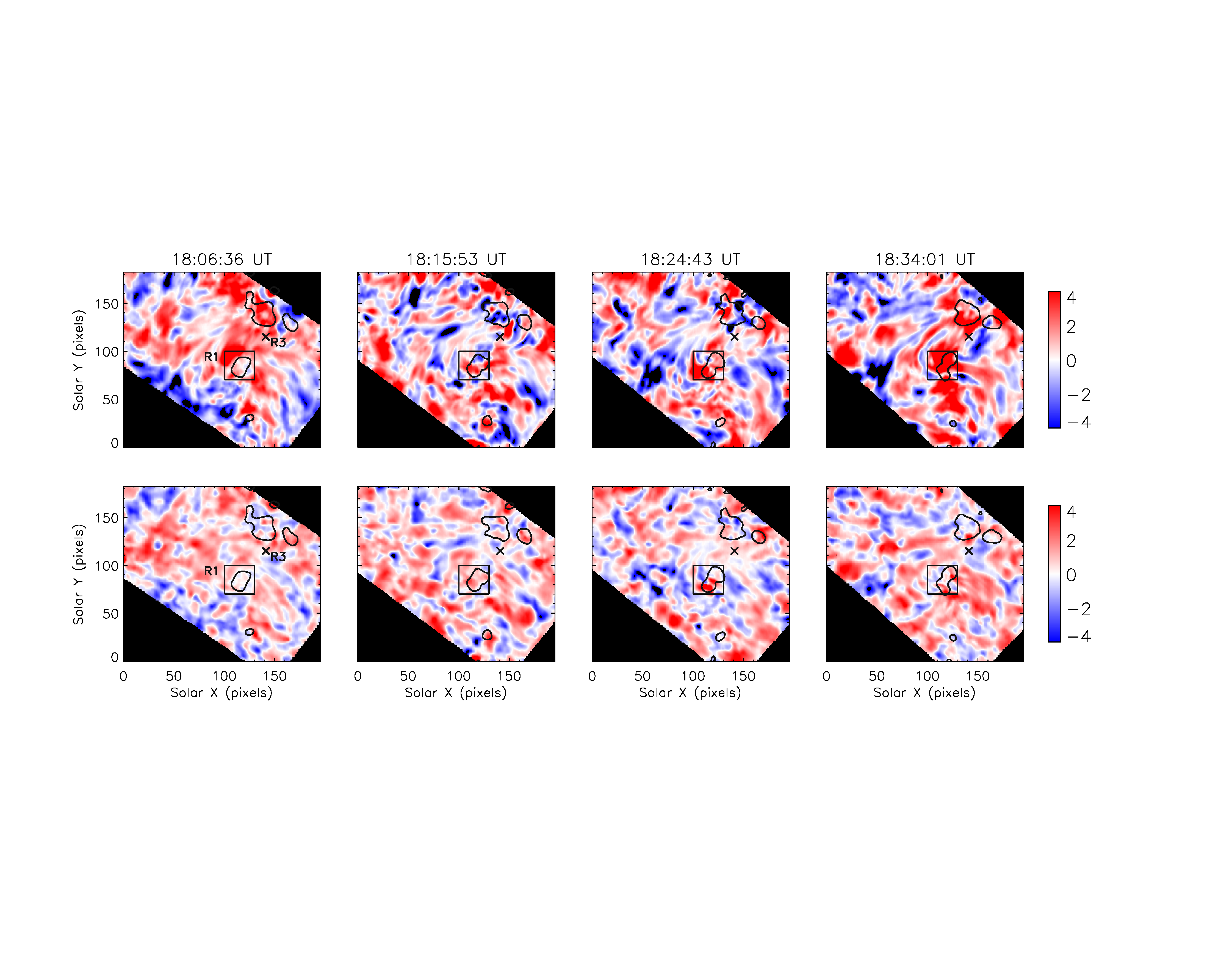}
\vspace{-3.5cm}
\caption{\ha\ (top row) and \ca\  (bottom row) Doppler velocity images at the corresponding times as in Fig.~\ref{fig2}. The square indicates where the Doppler-shift temporal variation in the south footpoint  was obtained. The cross denotes the centre of the small area from which the Doppler-shift temporal variation in the \ha\ loop was obtained. The pixel size is 0.16\arcsec.}
\label{fig4}
\end{figure*}

 The top and bottom panels of Fig.~\ref{fig4} show the Doppler-shift images in the \ha\ and \ca\ lines, respectively, obtained through the lambdameter method. We note that the Doppler shifts are calculated relative to an average profile taken from the area denoted with the larger square on the first panel of the \ha\ images in Fig.~\ref{fig2}. Thus, the Doppler velocities are only indicative of the flow patterns. Corresponding animations of  the Doppler-shift  \ha\ and \ca\ temporal sequences are also provided in Figs.~\ref{figa1} and \ref{figa2}, respectively.  Despite the obviously complex  structure of this region, the Doppler  maps clearly show the pattern of the HLs that appear with elongated blue- or red-shifted emission. Predominant downflows are observed in the legs of the CBP loops, but upflows with an elongated shape associated with the CBP structure are also present. Mottles are also easy to detect propagating in all directions forming the so-called rosettes \citep[see section~4.2.1 in][]{2012SSRv..169..181T}.

\begin{figure}[!ht]
\center
\hspace{-.6cm}
\includegraphics[scale=0.5]{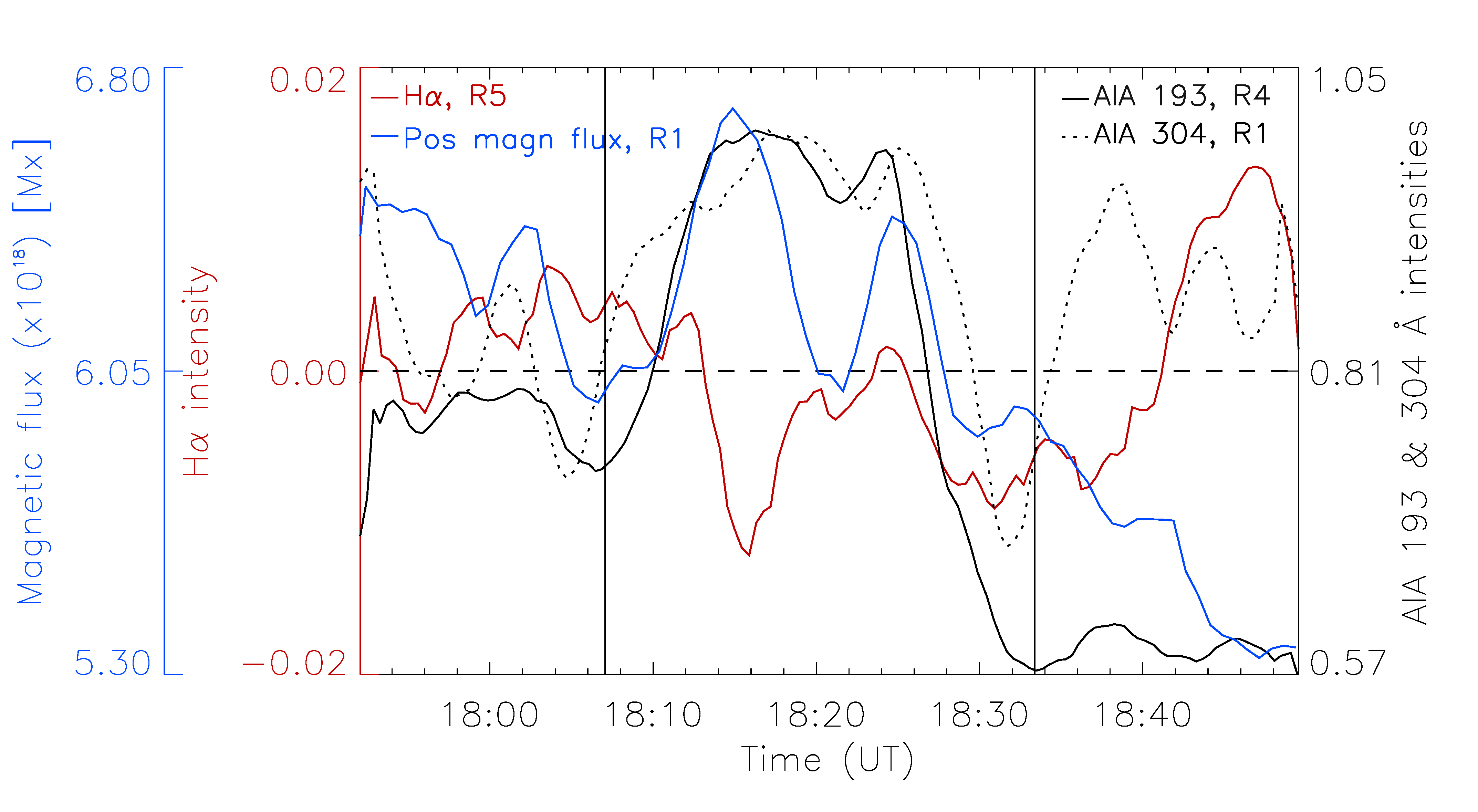}\\
\includegraphics[scale=0.5]{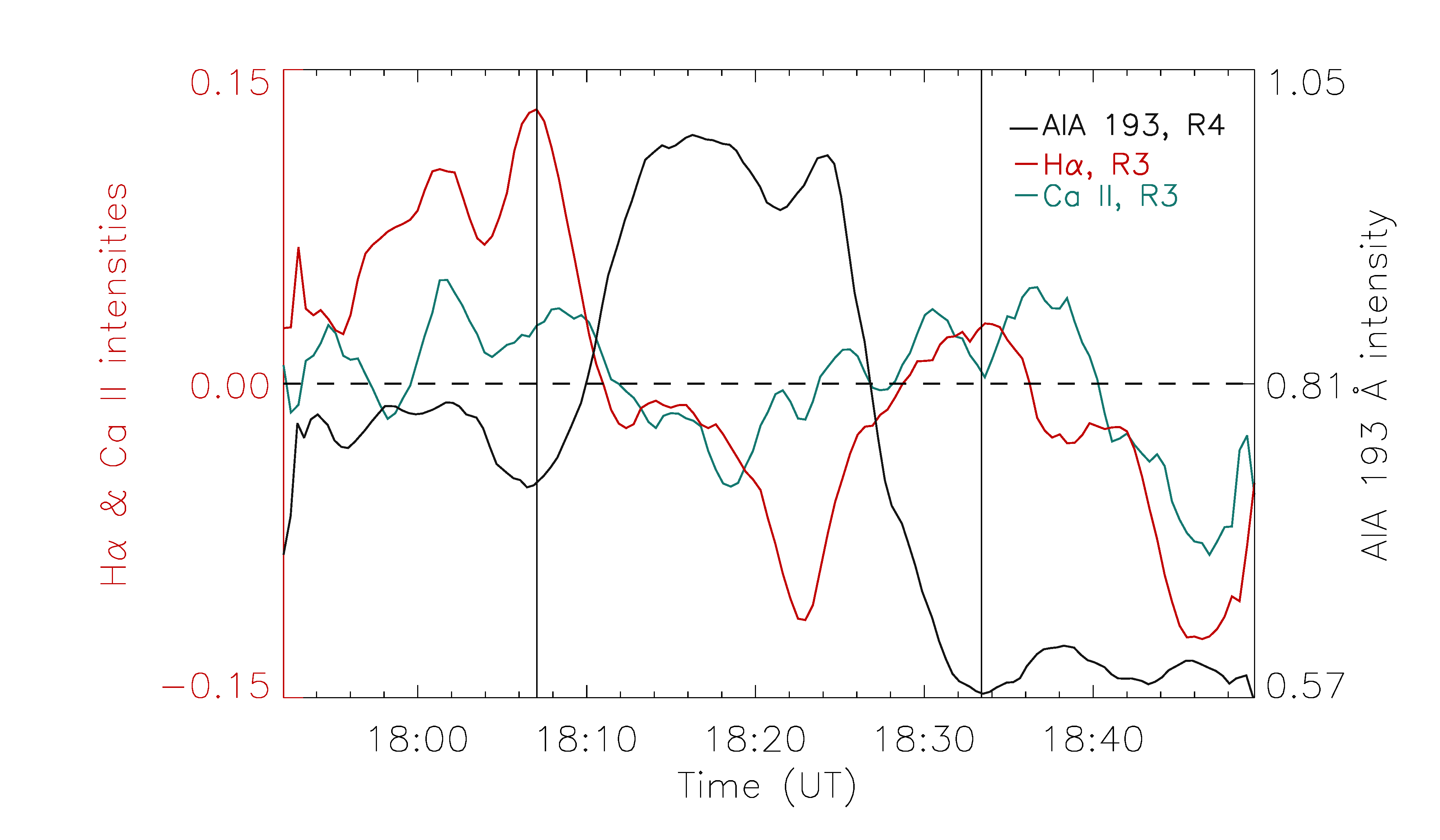}
\includegraphics[scale=0.5]{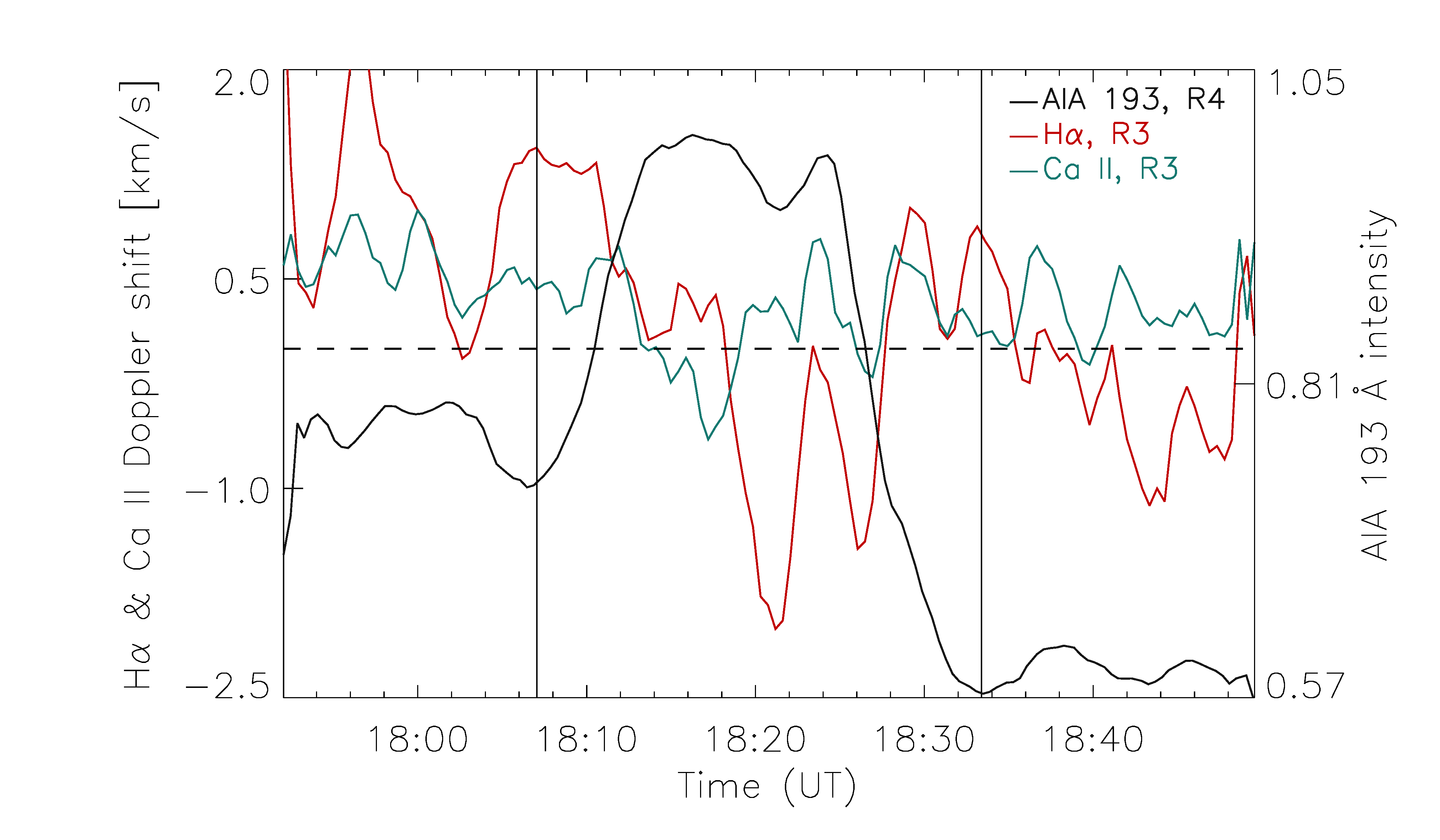}
\caption{Temporal variations of the coronal, transition-region, and chromospheric emission, \ca\ and \ha\  Doppler shift, and positive magnetic flux of the CBP. Top panel: The total positive magnetic flux and \aac\ normalized  intensity taken from R1 shown in Fig.~\ref{fig2},   \aa\ normalized intensity produced from R4 shown on the \aa\ in Fig.~\ref{fig2}, and the \ha\  intensity produced from the sum of the intensity in the blue and red wing at $\pm$45~\AA\ averaged from R5. Middle panel:  The \ca\ and \ha\ line-core intensities  produced from  one of the HLs, R3. Bottom panel:  \ca\ and \ha\ Doppler-shift variations in R3  as in the middle panel. The horizontal dashed line indicates the zero Doppler-shift value.  The \aa\ intensity temporal variations  (R4) are  shown for reference in both the middle and bottom panels.  }
\label{fig5}
\end{figure}

\begin{figure}[!ht]
\center
\hspace{-0.6cm}
\includegraphics[scale=0.50]{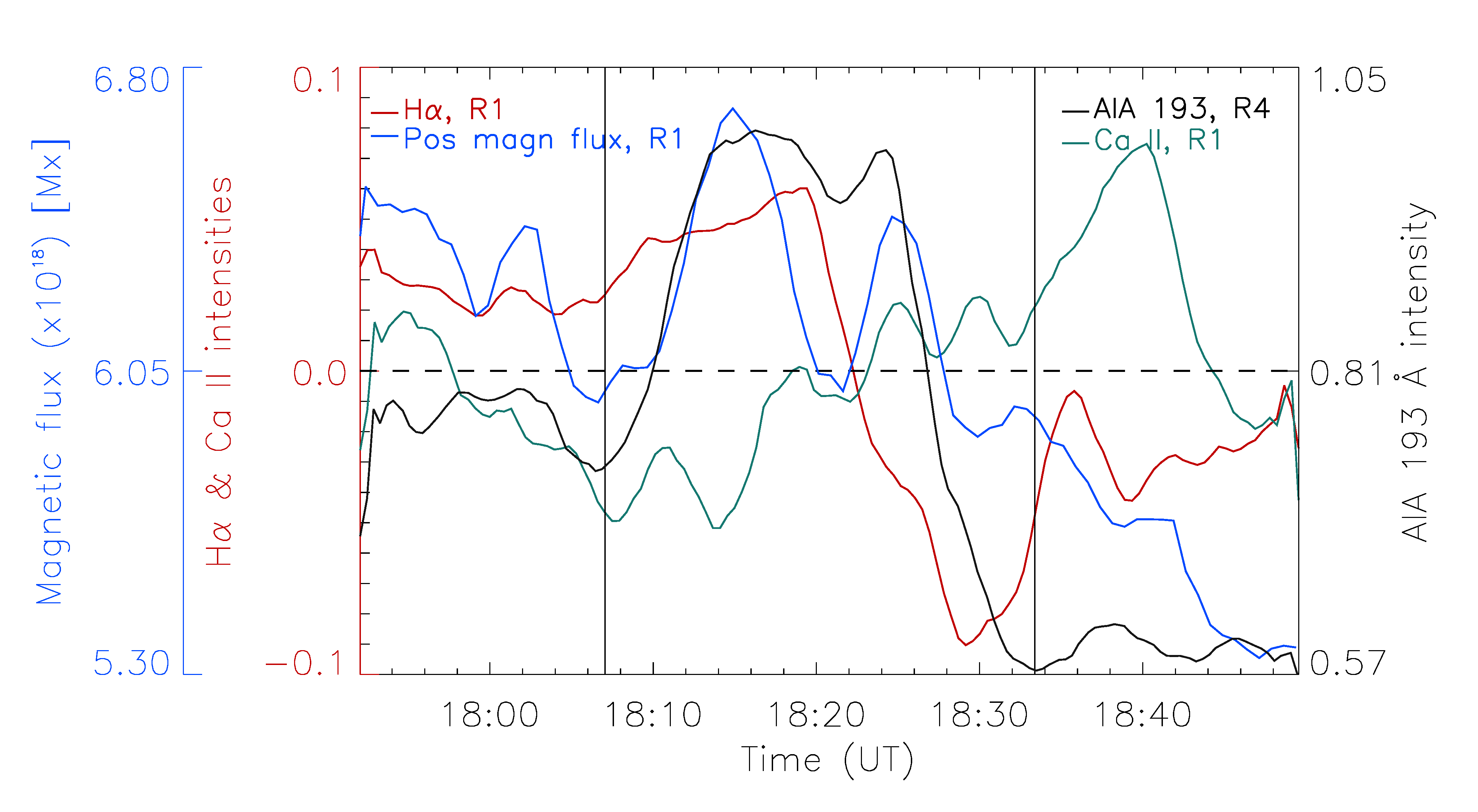}\\
\includegraphics[scale=0.50]{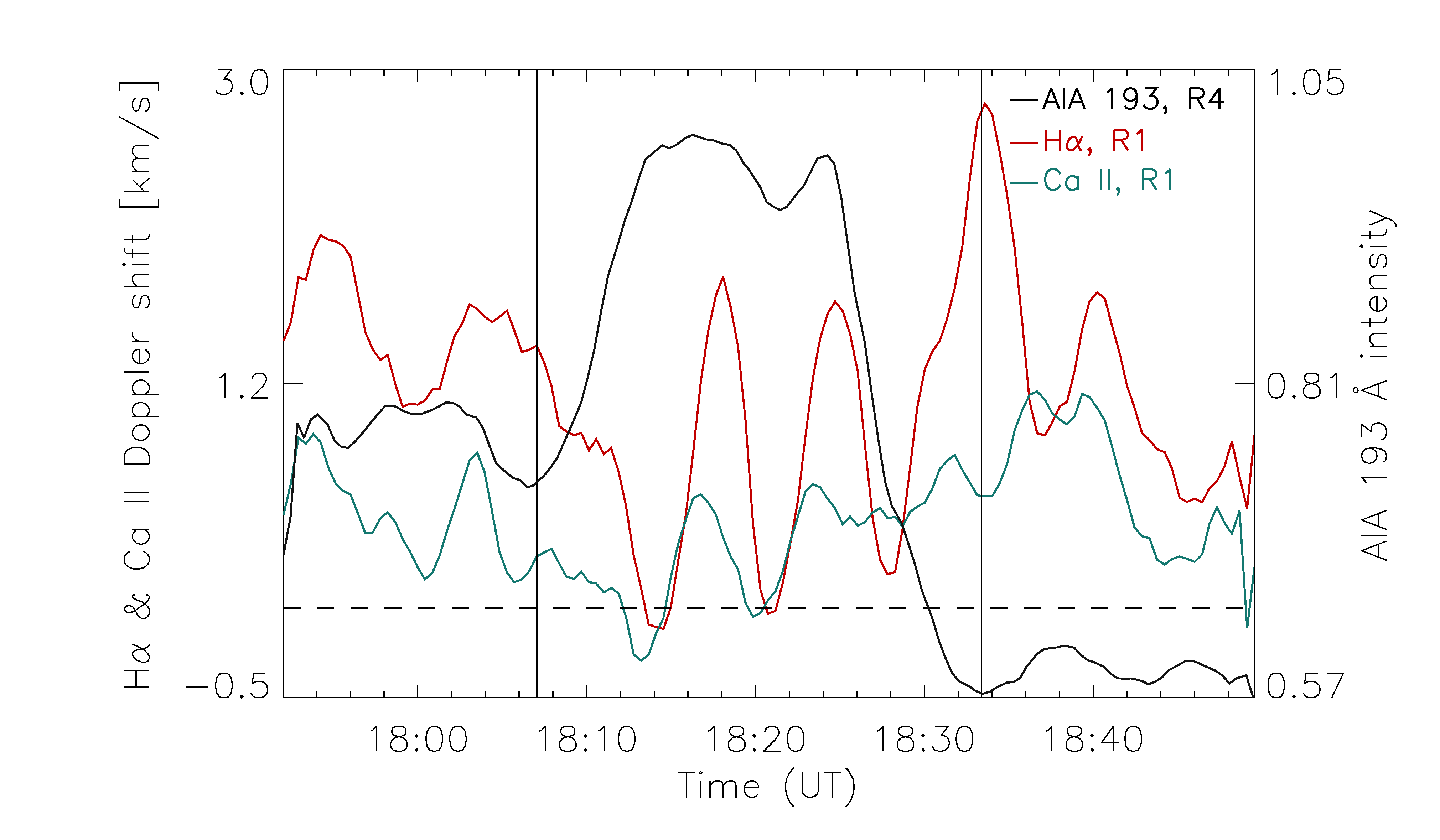}
\caption{ Temporal variations of the \ca\ and \ha\ line-core  intensities and \ca\ and \ha\ Doppler shift produced from the south footpoint of the CBP. Top: The total positive magnetic flux, \ca\ and \ha\ line-core intensities taken from  R1 together with the \aa\ intensity from R4. Bottom:  The \ca\ and \ha\ Doppler-shift taken from  R1, and  the \aa\ intensity from  R4.  The horizontal dashed line indicates the zero Doppler-shift value.}
\label{fig6}
\end{figure}

\begin{figure}[!ht]
\hspace{-0.8cm}
\includegraphics[scale=0.50]{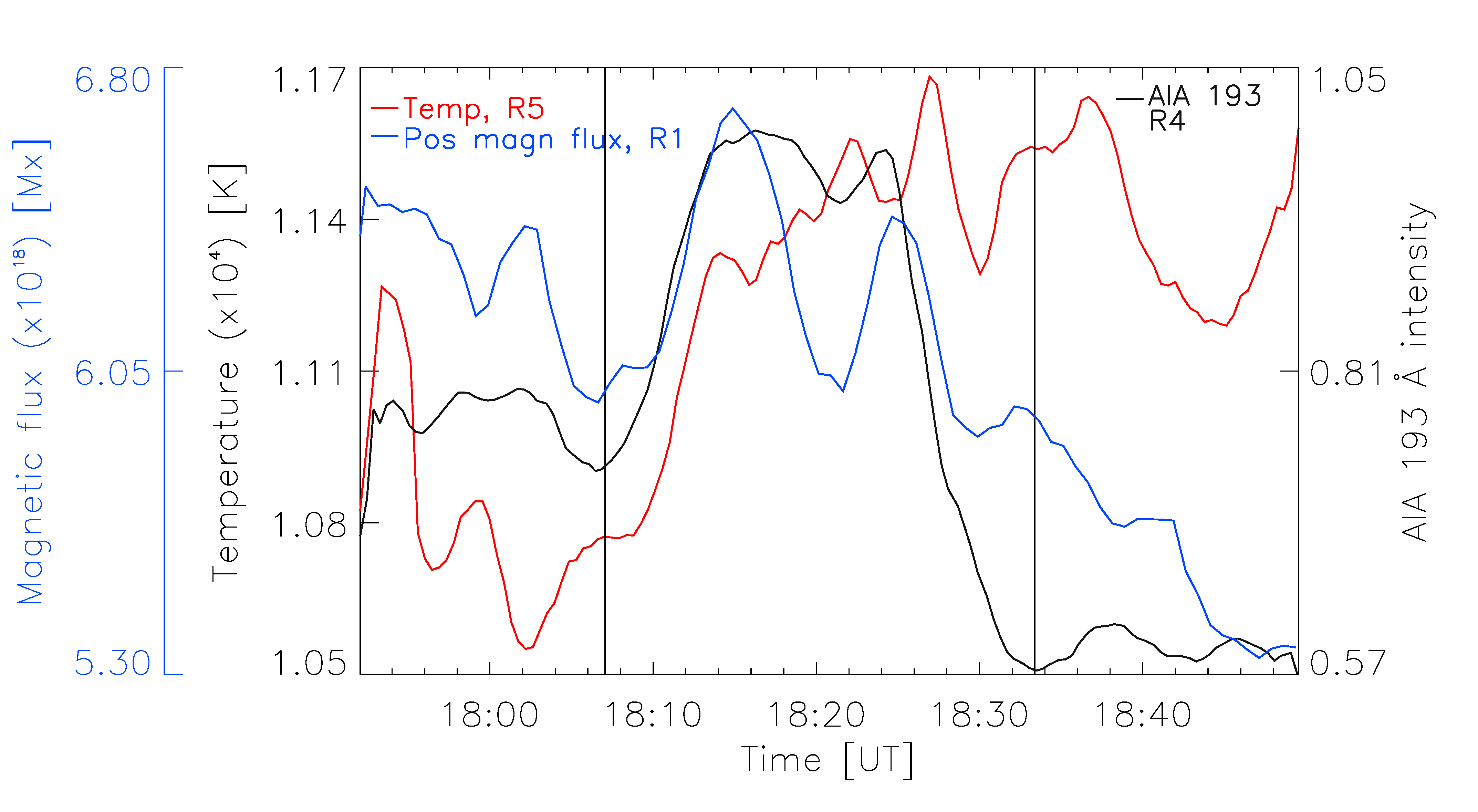}\\
\includegraphics[scale=0.50]{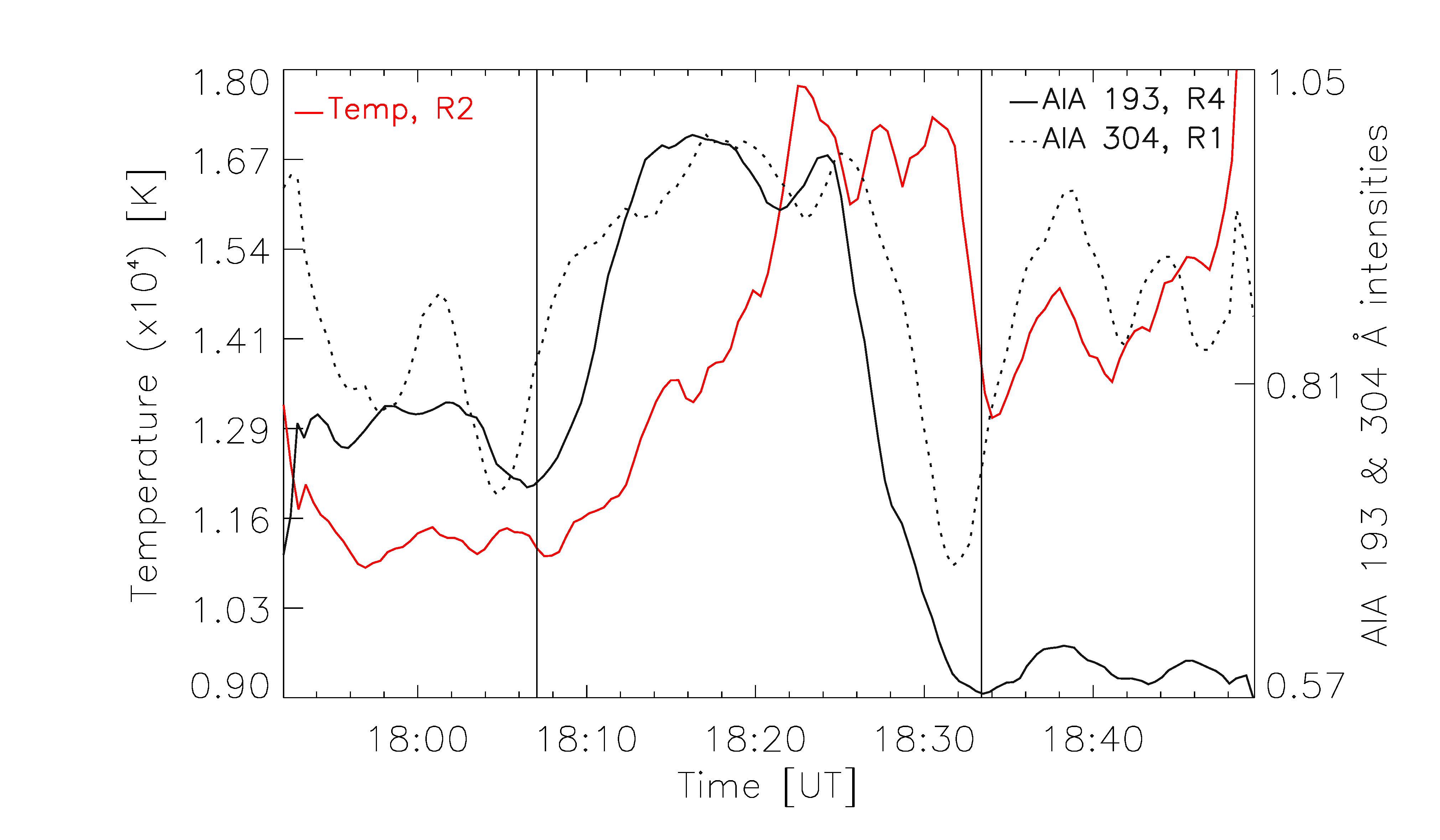}\\
\includegraphics[scale=0.50]{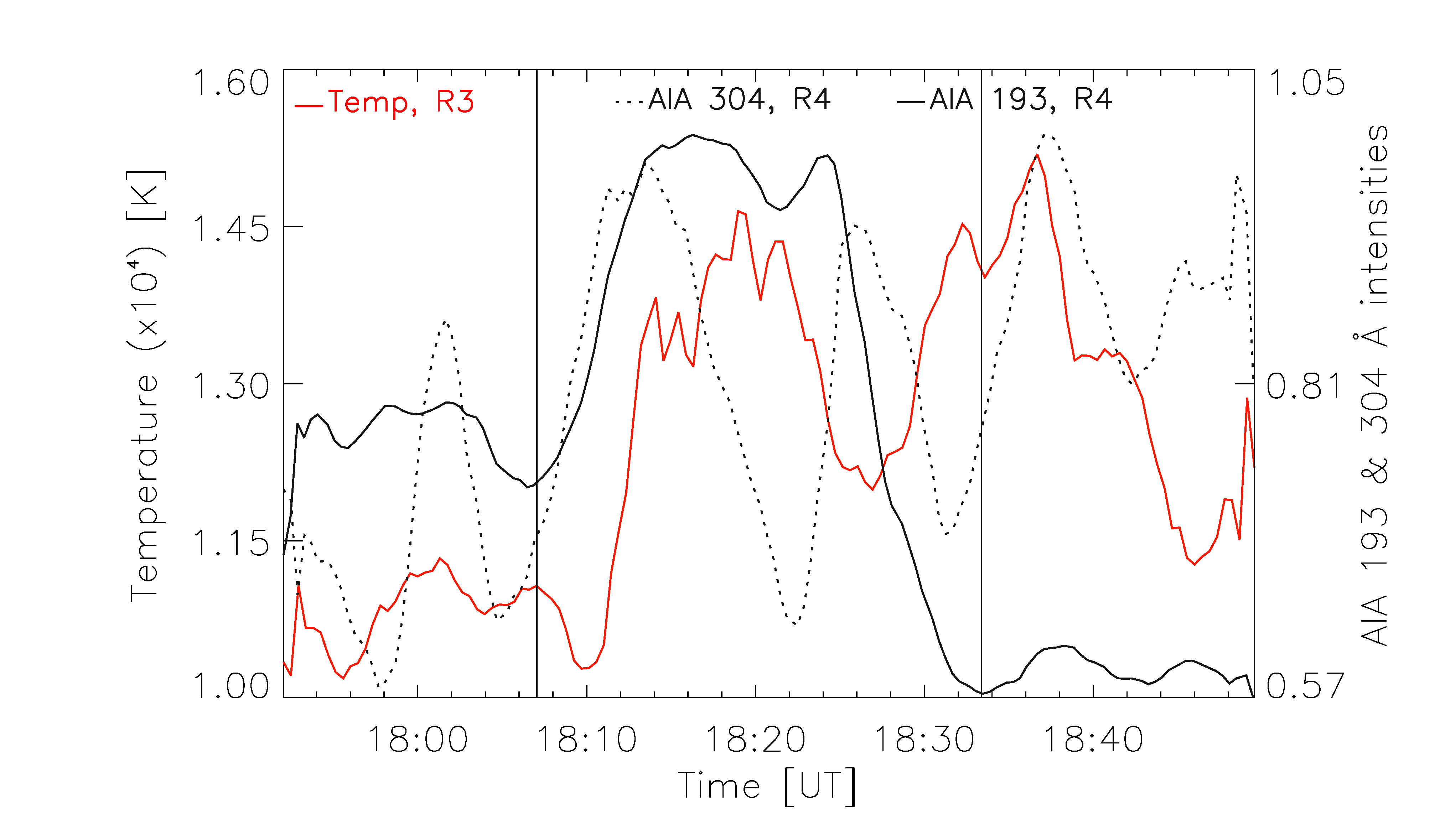}
\caption{Hydrogen temperature variations with respect to the magnetic flux, \aa,\ and \aac\ intensity variations of the CBP.  Top: The averaged temperature obtained from the whole CBP (R5). The total positive magnetic flux (R1) and the \aa\ intensity (R4). Middle: The averaged temperature in the CBP south footpoint, R2. The \aac\ is taken from the same area as the positive flux (R1). Bottom: The temperature estimated from the small area of one of the HLs, R3. The \aac\ is taken from the same area of the CBP loops as \aa, R4. }
\label{fig7}
\end{figure}

\begin{figure*}[!ht]
\centering
\vspace{-4cm}
\includegraphics[scale=0.9]{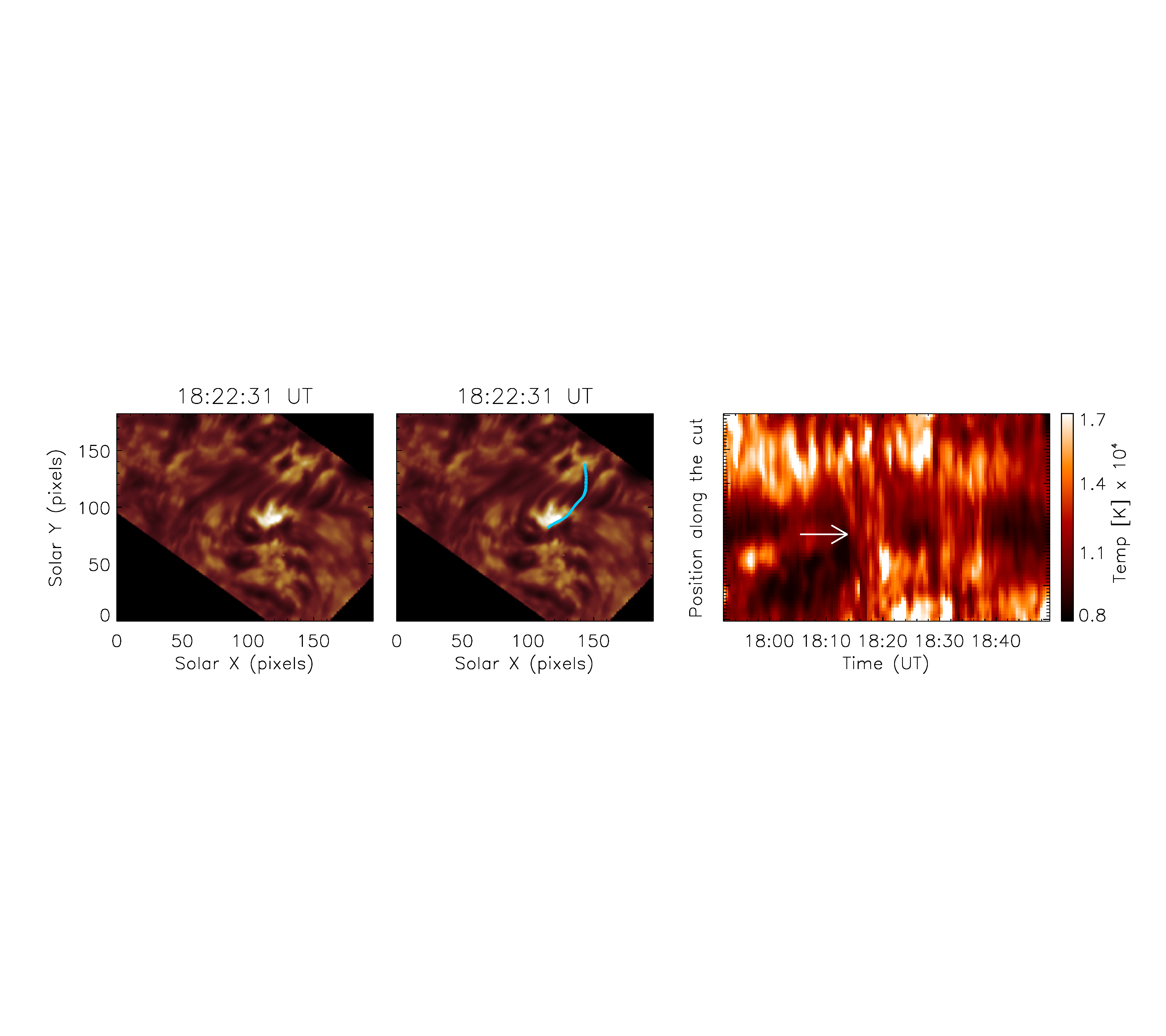}
\vspace{-5cm}
\caption{Temperature increase in the HLs. Left and middle: The \ha\ line-core intensity images. The blue line on the middle-panel image outlines the position of the pixels used to produce the hydrogen temperature time-distance image. Right: A hydrogen temperature time-distance image. The arrow points at the start of the heating episode of the HLs. The pixel size is 0.16\arcsec.}
\label{fig8}
\end{figure*}

\subsection{Chromospheric and coronal response to magnetic flux increase}
\label{heating}

\subsubsection{Intensity and Doppler-shift temporal evolution}
Although the CBP appears relatively unchangeable visually in the \aa\ channel during the FISS observing period, a short-lived brightness increase made us inspect the coronal intensity  in more detail. We derived the intensity at the chromospheric (\ha), transition-region (\aac), and coronal (\aa) temperatures, magnetic flux,  Doppler-shift, and hydrogen-temperature temporal variations from selected regions of interest (R) of the CBP including  the south footpoint of the CBP  (denoted as R1 and R2 in Fig.~\ref{fig2}), an individual HL (R3), the CBP loops (R4), and the whole CBP (R5 in Fig.~\ref{fig3}). 

 The top panel of Fig.~\ref{fig5} shows the total magnetic flux variations
 taken from the region of the positive flux outlined with a larger square
  (R1)  in the top row of Fig.~\ref{fig2}. We chose only the positive
 magnetic flux associated with the CBP because  it is centrally located  in  the
 FISS FOV and almost exclusively relates to the CBP, that is it provides  a less
 ``contaminated'' signal.  The negative magnetic flux  at the location of the
 CBP north footpoint is more dispersed and has more complex connectivities
 that are not related to the CBP. The top panel of Fig.~\ref{fig5}  also
 presents  the \aa\ averaged intensity variations
 taken from an area of the CBP coronal loops  (R4). We did not select the
 entire CBP because the \aa\ channel is also contaminated by
 transition-region emission. The inclusion of the CBP footpoints, and,
 especially, the south footpoint  of the CBP that is not obscured by mottles
 and cool-plasma upflows and strongly emit at low temperatures
 \citep{2010A&A...521A..21O},  may contaminate the extracted coronal
 emission. We also  present  the
   \aac\ temporal variations taken from the south footpoint (R1).

 Starting at $\sim$18:07~UT, a magnetic-flux increase (Fig.~\ref{fig5}, top panel) is detected reaching a maximum after
 $\sim$8~min.  A second but weaker peak follows $\sim$10~min
 later. For the first peak, the magnetic flux increase is
 $\sim$11\%\ with respect to the pre-event value ($\sim$6.0 $\times$
 10$^{18}$~Mx) which corresponds to a total magnetic-flux increase
 taken from an area of 31 $\times$ 31~px$^2$, of $\sim$6.7$\times$
 10$^{17}$~Mx for the first peak, and $\sim$6.4$\times$ 10$^{17}$~Mx
 for the second at $\sim$18:25~UT. The magnetic flux has been
 identified as a magnetic-flux emergence event through the co-temporal
 increase of the negative flux where the HLs are rooted. At
 $\sim$18:30~UT the magnetic flux drops to the values it had prior to
 the event. A co-temporal coronal intensity rise is detected in the
 coronal loops registered in \aa\ starting at $\sim$18:07~UT. 
   Surprisingly, the intensity increase in the \aac\ channel clearly precedes
   both the magnetic flux and the \aa\ intensity by 
   a bit more 
   than 1~min. The emission also persists a minute longer after the \aa\ in
   the coronal loops. The most plausible explanation is that \aac\ depicts
   the effects of the magnetic flux emergence  earlier than the HMI and only
   when the magnetic flux reaches above certain threshold of detection it is
   recorded by HMI. The same behaviour of the \aac\ emission is seen around
   18:00~UT where a small increase of the magnetic flux  again lags
   behind the transition region emission. This explains also the surprising
   lack of delay between the magnetic flux emergence and the coronal response,
   which is commonly observed during CBP formation \citep[for details see
     table~1, last column in][]{2018A&A...619A..55M}. Follow-up studies
   hopefully will shed more light on the behaviour of the \aac\ emission in CBPs.
 
 We should note that the HMI cadence is 45~s, and, therefore, the data
 are not fully co-temporal.  We prepared image sequences of the \aa, HMI, and
 FISS data, which are as close to each other in time as possible; see the
 animation in Fig.~\ref{figa1}. The clearly identifiable period of a
 coronal-intensity increase is roughly marked by two vertical lines (added to
 all curve figures in this paper) and has been determined from the coronal
 (\aa) intensity flux variations of the CBP.  The \aa\ intensity rise is
 approximately 35\%\ with respect to the pre-event value (at $\sim$18:07~UT). The maximum
 intensity is reached in $\sim$9~min at $\sim$18:16~UT and it remains
 relatively constant with a small dip before a second increase apparently
 related to the second magnetic-flux intensification.  We therefore conclude
 that an enhancement of the magnetic flux due to magnetic-flux emergence
 observed at photospheric level is associated with a heating of the CBP loop
 plasma at coronal temperatures.  Checking with the long-term evolutionary
 history of this CBP (see Fig.~\ref{fig1}), we see that, after the end of
 this heating episode, the CBP continued its evolution by becoming smaller
 and fainter and disappearing some 4~hrs later. 
 
 The \ca\ line-core
   intensity in the HLs and south footpoint varies differently from the
   \ha\ line-core intensity (for details see Figs.~\ref{figa1} and
   \ref{figa2}) most likely for two reasons. The HL are optically
   thin in the \ca\ line so that the behaviour of the \ca\ light curve is not
   comparable to the \ha\ light curve. Another reason is that the
   \ca\ light curve is highly ``polluted'' by the emission from the so-called
   bright grains \citep[for more see the review by][]{1991SoPh..134...15R},
   which have a marked presence in the \ca\ line but not in the
   \ha\ line. These bright grains are produced from shocks generated by
   vertically propagating acoustic waves that are just above the acoustic
   cut-off frequency, emanating from the photosphere
   \citep{1997ApJ...481..500C}. However, it should be noted that the
   downflows in the footpoints are well distinguishable in both lines; for
   details see the results of the Doppler velocities later in this section.
 
 Next, we investigated the response of the chromospheric counterpart of the
 CBP to the magnetic-flux increase. First, we noticed an enhancement of the
 chromospheric activity associated with both the occurrence of HLs  and
 mottles following the magnetic flux increase (see the animation in
 Fig.~\ref{figa1}). This is shown with  an \ha\ light curve in the top panel of Fig.~\ref{fig5} produced from an area  enclosing most of  the CBP (R5 in Fig.~\ref{fig3}). The light curve is obtained from the sum of the intensity in  the blue and red wings of \ha\ at $\pm$0.45~\AA, after subtracting and normalizing by  the mean intensity.  There is a clear decrease (enhanced presence of absorption features) with respect to the average intensity that indicates enhanced dynamics associated with new mottles and HLs.  It should be noted that this activity appears to last longer (until $\sim$18:40~UT) than the heating event in the corona. As it could be seen later from the temperature analysis (Section~\ref{temp}), a hydrogen temperature enhancement also continues after the coronal response has ended. This is also visually noticeable from the middle and right panels of Fig.~\ref{fig3} and  the provided animations in Figs.~\ref{figa1} and \ref{figa2}.  

More detailed analysis was made for one of the HLs. In the middle and bottom panels of Fig.~\ref{fig5} we show the \ha\ line-core intensity  and Doppler-shift temporal variations, respectively,   obtained from a small 3 $\times$ 3 px$^2$ area (R3, see the cross sign on  the \ha\ images of Fig.~\ref{fig2} and Fig.~\ref{fig3}) at one of the HLs. The reference profile was chosen as an average over the entire dataset from a region outside the CBP as shown in Fig.~\ref{fig2} (first panel in the third row,  `Ref').    The \ha\ line-core intensity curve is produced by  subtracting and normalizing by  the mean \ha\ line-core intensity of this area.  A signature of the HL is notable in the \ha\ curve after $\sim$18:16~UT (Fig.~\ref{fig5}, middle panel).
 The HL disappears at $\sim$18:28~UT, which gives a life time of $\sim$12~min. However, this life time should be taken with caution because it looks like the HL  has moved horizontally, thus leaving the inspected location. Taking a larger  area would have resulted in including another  feature. As seen from the right panel in Fig.~\ref{fig3}, the HL seems to be present  for longer (i.e. 27 min).  After inspecting the animation, we concluded that a new upflow  from the CBP footpoint along the same field line may possibly have led to the reappearance of this HL in the \ha\ images. The Doppler-shift pattern of the HL  (Fig. \ref{fig5}, bottom panel) indicates upflows followed by downflows. It should be noted that as we investigated a location at almost the horizontal part of the loop, the Doppler values of the up- and downflows are very small. We inspected the animation sequences, which indicate that the \ha\  plasma in the HL  seems to originate from upflows  from  the footpoints of the CBPs (see the animation in  Fig. \ref{figa1}).  
 
 We also investigated the general pattern of the \ha\ line-core intensity (produced after subtracting and normalizing by  the mean  \ha\ line-core intensity of this area) and Doppler shifts in the south footpoint of the CBP
 (Fig.~\ref{fig6}, top and bottom panels, respectively). To accomplish this, we selected the same area as for the magnetic
 flux analysis  (R1).  The \ha\ line-core intensity curve indicates an
 intensity increase in the south footpoint, which begins together with the
 magnetic-flux increase. Almost 10~min after the start of the coronal
 heating period a distinct pattern of up- and downflows is apparent from the \ha\
 Doppler-shift curve profile (Fig.~\ref{fig6}, bottom panel), indicating increased chromospheric activity that is also
 discernible in the Doppler-shift curve of the \ca\ line, although weaker. After 18:28~UT, a
 steep \ha\ line-core intensity decrease (Fig.~\ref{fig6}, top panel) is associated with strong up and
 down cool-plasma flows that appear to be related to large
 mottle or surge-like events not related to the CBP phenomenon. The features
 propagate east and west of the CBP originating from the same polarity as
 the CBP south footpoint.

\subsubsection{Temperature evolution}
\label{temp}

 Important information is brought by the obtained hydrogen temperature at chromospheric levels which, as a reminder,  was obtained with the multi-layer inversion technique of \citet{2020A&A...640A..45C}, as explained in section~\ref{obs_met}. In Fig.~\ref{fig7} we show the   temperature  temporal variation  averaged over the whole CBP (R5  shown in Fig.~\ref{fig5}, see also the animation in Fig.~\ref{figa3}).  For reference we also show the \aa\ and positive magnetic-flux temporal variations. The temperature of the whole CBP  starts to rise steeply  with a delay of less than $\sim$3~min (at $\sim$18:10~UT) following the magnetic-flux and the coronal emission enhancements. The average temperature increase is $\sim$10\% with respect to the pre-event temperature, from 1.07 $\times$ 10$^4$~K  to 1.175 $\times$ 10$^4$~K. The maximum value is reached $\sim$17~min later at 18:27~UT. The random temperature  error in each pixel was estimated \citep{2020A&A...640A..45C} at 140~K. The chromospheric heating event lasts longer than the coronal heating event. 
 
We also derived the averaged temperature from a small area of 11 $\times$ 16~px$^2$ above the location of the positive flux in the south footpoint of the CBP  indicated with a small black square  on the HMI magnetograms in Fig.~\ref{fig2} (R2).  Figure~\ref{fig7} (middle panel) shows the clear pattern of a significant  temperature increase in the south footpoint of the CBP that start with a delay of less than $\sim$3~min after the coronal intensity enhancement. The temperature initially increases reaching a small peak (at 1.35 $\times$10 $^4$~K) co-temporally with the magnetic flux and coronal intensity peaks. Initially, the increase is around 17\%, from 1.15 $\times$ 10$^4$~K  at  $\sim$18:10~UT to 1.35 $\times$ 10$^4$~K at $\sim$18:15~UT. For the following 13~min the temperature sharply continues to rise, reaching  $\sim$1.78 $\times$ 10$^4$~K at $\sim$18:23~UT. Thus the increase with respect to the pre-event temperature is 55\% (from 1.15 $\times$ 10$^4$~K to 1.78 $\times$ 10$^4$~K). We believe that the longer duration  of the temperature enhancement is related to the enhanced activity of downflows, which is described Sect.~\ref{disc}.  The temperature falls  back sharply  more than $\sim$20~min later after the beginning of the coronal emission decrease has begun. 

The temperature curve from one of  the HLs (R3) was also produced and is shown in the bottom panel of Fig.~\ref{fig7}. Again, the temperature increase in the HL location is evident. This increase starts with the appearance of the HL and lasts until the HL vanishes (see also the middle panel of Fig.~\ref{fig5}). The temperature increase is 46\%, from 1.03 $\times$ 10$^4$~K to 1.5 $\times$ 10$^4$~K. The maximum is reached during the re-appearance of the HL at 18:37~UT.  The \aac\  and the \aa\  light curves both taken from the CBP loops show a co-temporal increase, while the emission in the \aac\ (i.e. cooler transition-region loops) lasts a few minutes longer.

Furthermore,  we  demonstrate  the CBP  temperature increase with a  time-slice image shown in Fig.~\ref{fig8}. It was produced from the location of one of the HLs  discussed above and traced in the \ha\ line-core image taken at 18:21:31~UT (shown in the left and middle panels in Fig.~\ref{fig8}).   This figure clearly shows the heating period of the HLs. 

We should mention again (see section~\ref{intro}) the study by \citet{2009A&A...507.1625C}  reporting on the morphological and dynamical evolution of an AFS observed in the \ca~8542.1~\AA\ line. We established that this AFS also represents a chromospheric counterpart of a CBP. The data were obtained in 2006 and therefore the only available coronal  imaging data are from SoHO/EIT. The IBIS data were taken during the late stage of the CBP evolution called the decay phase. As in our study, in addition to the AFS rooted in a bipolar region, mottles showing upward and downward motions were also observed to evolve from these magnetic-flux concentrations. The study estimates the temperature of the AFS at 8.2 $\times$ 10$^4$~K  in one footpoint and at about 2.4 $\times$ 10$^4$~K in the other,  based on the \ca\ Doppler width (see their eq.~2). These temperatures are higher than the footpoint values of 1.78 $\times$ 10$^4$~K obtained in the present study.  The study  also  reports upflows in the body of the AFS and downflows in its footpoints during 7~min of observations (see their table~2). The Doppler pattern  is consistent with  that obtained in this work during certain time periods. Our analysis identifies more complex evolution that may also be related to the magnetic-flux increase in our CBP.

 \subsection{CBP magnetic topology}
\label{topology} 

\subsubsection{The overlying corona}\label{overlying_corona}

We also investigated the magnetic configuration of both the
  low-lying CBP loops and the overlying corona. Starting with the
  latter, we calculated a simple potential field extrapolation from
  the HMI magnetograms for four observed times, namely one at 17:21~UT, that is 
  before the flux emergence and heating event described in the
  foregoing sections (section~\ref{heating}), and three during the event,
  namely at 18:10 UT, 18:15 UT, and 18:25 UT. The calculated 3D
  structure was then searched for the presence of null points. To
visualize and analyse the 3D magnetic field the Visualization and
Analysis Platform for Ocean, atmosphere, and solar Researchers (VAPOR)
software was used.


\begin{figure}
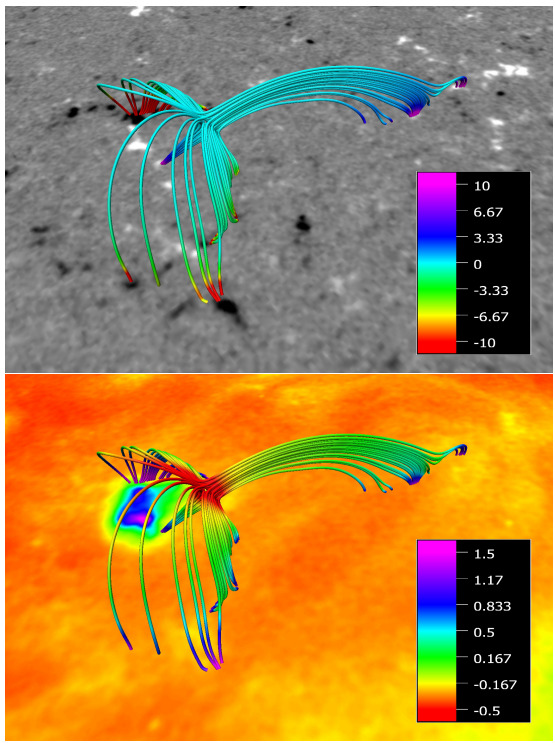

\ifnum \inclfigs >0
\hbox{
\includegraphics[width=0.8\columnwidth]{fig9a_39329.png}}
\hbox{
\includegraphics[width=0.8\columnwidth]{fig9b_39329.png}}
\fi
\caption{Potential field extrapolation of the coronal magnetic
  field overlying the CBP at 17:21~UT. Top panel:  The background image is the HMI
  magnetogram from which the extrapolation has been made. The field lines are coloured
  according to the value of the vertical field component, with values in
  gauss as in the colour bar. 
  Bottom panel: The same but overlaid on a co-temporal \aa\ image. The field
  line colours correspond to the logarithm of $|{\bf B}|$.}
\label{fig9}
\end{figure}

\begin{figure}[!ht]
\hspace{-0.8cm}
\ifnum\inclfigs>0
\includegraphics[scale=0.65]{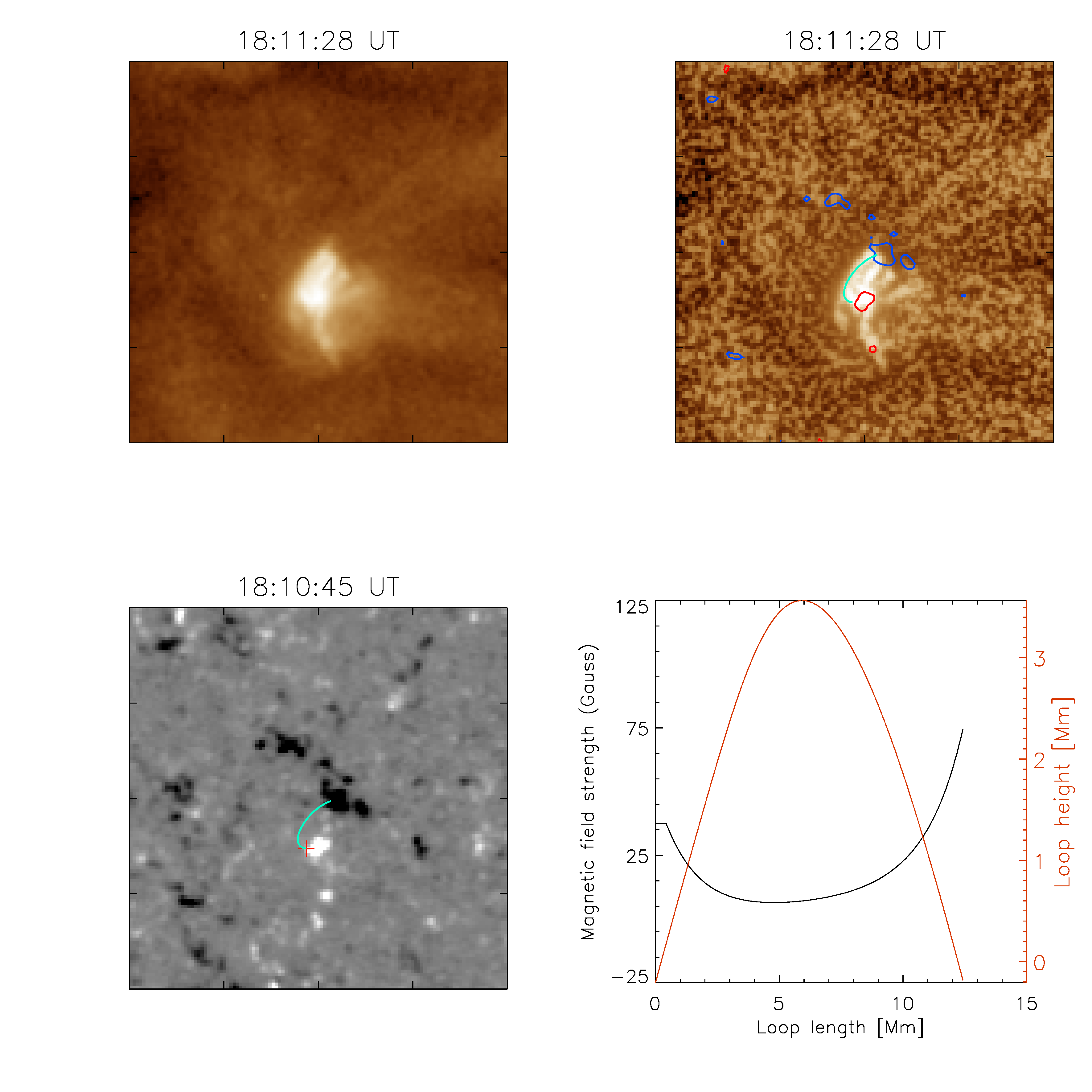}
\fi
\caption{Linear force-free field extrapolation of a CBP loop. Top left: \aa\ image. Top right: A MGN processed \aa\ image with overplotted a  magnetic field line that matches one of the extrapolated CBP loops. Bottom left: An HMI magnetogram with the same magnetic field line. The cross indicates the location from which the magnetic field line was traced. Bottom right: Magnetic-field-line  parameters.}
\label{fig10}
\end{figure}

For the extrapolation, HMI data with a large FOV of
  $349.5\arcsec \times 349.5\arcsec$ were chosen as a basis. This is much larger than the area shown for the \aa\ observations
  in Fig.~\ref{fig2}  ($21\arcsec  \times 39\arcsec$), amounting to a total of 694 px $\times$ 694
  px with resolution $0.5\arcsec$~px$^{-1}$.  The base area was
  essentially in flux balance, where the deviation between the total
  positive and negative flux is at the $2\%$ level. To prevent boundary
  effects, before extrapolating the base data were apodized in a band
  of $40 \arcsec$ at the rim, maintaining their average. We checked
  that the field-line geometry of the area of interest shown below is
  the same irrespective of the apodization.

Figure~\ref{fig9} illustrates the field-line structure above the CBP
for the 17:21 UT observation. To
facilitate the visualization, we made the choice of field lines giving extra weight to those that pass near to weak-field locations, so
as to better illustrate the region around possible null points.  We
see that the field configuration is  the fan-spine kind, in which the fan
surface has an inclined dome shape that reaches down to the
negative polarity at the surface (to help with the orientation: the
string of negative elements towards the top-left of the top panel are
the forebears of the set of negative elements in the top row in
Fig.~\ref{fig2}, which are for later times). Like in the simplest
fan-spine configuration, the lower spine is rooted in one of the
elements of the parasitic positive polarity. The parasitic
polarity is distributed in a string of magnetic elements. The lower
spine shown in the figure just attaches to one of them; the field
lines issuing from the other elements are loops contained within the
space below the fan surface.  The upper spine, in turn, does not reach
high in the corona but, instead, bends over and connects to positive
elements outside of the CBP region to the right in the figure. In the
bottom panel, we see that the CBP as detected by \aa\ is located in a
fraction of the dome only, similar to the situation described by
\citet{2018A&A...619A..55M}.

There remains the question of whether there is a null point at the
junction of the spines with the fan surface. To search for this null point, we used the
method developed by \citet{2007PhPl...14h2107H}. Through this method, we located a prominent null point exactly at that juncture, at a height of $6.8$ Mm above the surface. Comments to the structure
and possible presence of null points for other observed times are
deferred to the discussion.

\subsubsection{Low-lying loops: LFFF extrapolation}

Next we investigated in detail the CBP magnetic field structures that
confine plasmas emitting at coronal temperatures (see
section~\ref{disc}) using the LFFF method as explained in
section~\ref{obs_met}. We determinted the value of $\alpha$L by
optimizing the match of the calculated loops to those apparent in the
MGN-processed \aa\ images.   For the LFFF extrapolation we used the higher
signal-to-noise HMI 12~min cadence data.  We started the loop
identification with \aa\ data taken as early as 16:00~UT, but failed
to obtain any LFFF field lines that match the observed loops. At this
time the CBP loops appeared strongly sigmoidal or curved, that is sheared,
implying the presence of large free magnetic energy.  Visual analysis
of the AIA time series showed an ejection of cool material just before
16:00~UT, similar to the cases discussed in
\citet{2018A&A...619A..55M}. This clearly suggests that an eruptive
phenomenon has taken place. As shown above (Fig.~\ref{fig1} and related
discussion), a magnetic-flux increase started as early as 14:00~UT,
which has possibly contributed to this activity.   
We were only able to identify some individual loops that could be fit with
the LFFF model after 17:12~UT. In Fig.~\ref{fig10} and Fig.~\ref{figa4} we give
examples of two extrapolated loops and their parameters.  We
concentrated our analysis to only the largest loops because the small loops
lie far too low in the solar atmosphere (below 1~Mm) and the
extrapolation method becomes less reliable. Low-lying loops, including
chromospheric loops, in the future will be modelled using the newly
developed magneto-hydro-static extrapolation model
\citep{2019A&A...631A.162Z}, which has higher accuracy for
extrapolations at chromospheric heights. The analysis established that
field lines with $\alpha$L of $-$3.0 and $-$3.5 best match the
identified \aa\ loops. During the analysed one-hour observations the
$\alpha$L parameter of the best-fitting field lines remained unchanged.
The maximum loop heights were $\sim$3.5~Mm and lengths $\sim$12~Mm,
respectively.  Higher and longer field lines failed to align with the
observed loops. The presented loops have magnetic field strengths of
75~G (negative polarity) and 25 and 35~G (positive polarity) at the
photosphere, for the loops shown in Fig.~\ref{fig10} and
Fig.~\ref{figa4}, respectively.  As found by a study of
\citet{2010ApJ...723L.185W} and recently confirmed by
\citet{2017ApJ...842...38X}, the magnetic field strength is typically
different in the two loop footpoints.

\section{Discussion}
\label{disc}

The heating of plasmas confined in small-scale loops has been explored
through numerous theoretical models \citep[for a review see section~9
  in][]{2019LRSP...16....2M}. A classic model for flux cancellation CBPs was
proposed by \citet{Priest_etal_1994} (see also
\citealt{Parnell_etal_1994_3d}). This was a highly idealized,
analytical, purely 
magnetic model of a CBP in two dimensions. This model starts with two unrelated
opposite magnetic poles at the surface that move towards each other and
considers a quasi-static approach, in which the field relaxes to a potential
configuration continuously along the process.  After reaching a critical
mutual distance (the interaction distance), an X-type null point is
created in the vertical axis between these two magnetic poles followed by reconnection 
at the X-type null point. The model was then tested numerically and extended by
\citet{Von_rekowski_etal_1_2006,Von_rekowski_etal_2_2006} by solving the
two-dimensional magnetohydrodynamics (MHD) equations for magnetic field and plasma in the
isothermal, gravity-free approximation. Another fundamental approach using
analytical theory, now in three dimensions, was made by
\citet{Longcope_1998}, who considered the approach of two magnetic poles at
the surface that are nearing each other but not exactly along the line
joining  them. A 3D numerical model of this sort of flyby process was
carried out by  \citet{Galsgaard_etal_2000}, who provided first glimpses of
the complication of the 3D reconnection patterns that may be at the basis of
some of the CBP structures.

The recent model by 
\citet{2018ApJ...864..165W} investigates the formation of CBPs in
coronal holes. The study also explores the formation of associated
collimated flows (i.e. jets), which is not the subject of the present
study and is not discussed further. In this model the initial
magnetic configuration corresponds to  a fan-spine topology with a single
3D null-point and with the null-point spine axis connecting to the
minority polarity flux.  The structure  was
stressed by three types of surface flows. The first was a large-scale surface flow 
shearing only one part of the separatrix surface, the second represents a large-scale surface flow
that shears part of the polarity inversion line around the minority flux, and for the third set-up the second shearing flow together with  a flyby of the majority polarity flux  that passes the moving minority polarity flux was employed. Small-loop structures (i.e. CBPs) with different morphologies were produced. The heating in these magnetic configurations
 was due to the shearing of the magnetic field near to the
separatrix, which leads to steady interchange reconnection. This is
coupled with quasi-periodic, low-intensity bursts of reconnection
resulting in CBPs with periodically varying intensity.  This model relies on a very high level of shear, far from the observational values, yet it shows that magnetic reconnection in the corona can
produce a phenomenon that resembles a CBP with varying intensity.

The magnetic field-line configuration obtained in
  Section~\ref{overlying_corona} for the overlying corona at 17:21 UT
  shows a clear fan-spine topology; the upper spine does not reach
  high coronal levels but, rather, bends over and connects to
  magnetic elements of positive polarity several tens of megameters away from
  the base of the CBP. By using the method of
  \citet{2007PhPl...14h2107H}, we also detected a clear null
  point at the junction between fan surface and spines. Interestingly,
  the projection of the \aa\ CBP observation fits very well in the area
  between the root of the lower spine in the parasitic positive
  polarity and the comparatively strong negative magnetic elements at
  the intersection of the fan surface with the photosphere on the
  northern side of the complex. This one-sidedness for the CBP with
  respect to the fan-spine configuration is also apparent in most of
  the field configurations for CBPs calculated by
  \citet{2017A&A...606A..46G}. It is precisely in that domain that the
  flux emergence and heating event described in detail in
  Section~\ref{heating} takes place. That heating event occurred some
  50~min after the 17:21~UT observation. To complete the coronal view,
  therefore, we also calculated potential field extrapolations
  for three times at the beginning (18:10~UT), in the middle (18:15~UT), and towards the end (18:25~UT) of said event. We obtained field-line configurations for all three times, which are very similar in
  their general fan-spine structure to that of the earlier time,
  albeit with the possibly important difference that for the first two, those at 18:10~UT and 18:15~UT, no null point could be
  detected. Those two times coincide with the rising period of the
  magnetic flux at the surface due to the flux emergence episode.

As already mentioned above, during the analysed observational period 
of only $\sim$1~h, restricted by the ground-based observations, we
identified a short period of photospheric magnetic flux increase in
the footpoints of the CBP related to flux emergence.  Aftereffects of
the magnetic-flux increase are observed in both the coronal and
chromospheric components of the CBP.  We detected a co-temporal rise
of the coronal emission, indicating that the  CBP loops
are heated to coronal temperatures. At the time of this coronal
emission enhancement the magnetic field concentrations where the
footpoint of the CBP are rooted do not show any significant
displacement or convergence. Therefore, if the heating has occurred at
coronal heights caused by magnetic reconnection, the driver is
possibly related to random small-scale footpoint motions that are
visually not possible to follow in the observations. If
 magnetic reconnection has taken place in the corona,
thermal conduction, which efficiently transports heat along
the magnetic field lines at temperatures above log T (K) $\ge$ 4.7,
would uniformly distribute the heating along the CBP loops. This would be detected
through  the coronal observations by an emission enhancement in
the coronal channels of AIA, for example \aa. The reconnection process must be relatively steady as no signature of bursty emission is detected in the coronal loops. Another possible explanation is that the reconnection occurred on spatial scales far smaller than the present instrumentation can spatially resolve, thus we detect only the aftereffects, namely the uniformly heated coronal loops along their entire length.

A heating episode is also detected in the chromospheric counterpart of
the CBP observed in the \ha\ line.  We make a small diversion here to
discuss the chromospheric morphology of the CBP. In the \ha\ line-core
intensity images the morphology of the CPB is seen as a bundle of elongated dark features
that have a sigmoidal shape at the one end (south).  As it has been
shown by \citet{2012ApJ...749..136L} from their 3D radiation-MHD
simulations combined with three-dimensional non-LTE radiative transfer
computations of the \ha\ line, dark elongated structures seen in the
\ha\ line core, referred to as fibrils, trace magnetic field lines to
a large extent. Their fibrils were explicitly found within $\beta$ =
0.01 contours, following the magnetic field direction between opposite
flux concentrations.  In the case of our CBP, these features connect
the CBP bipoles, and therefore represent the chromospheric counterpart
of the CBP. Thus, the term we adopted, HLs, best
describes the observed features.  The  HLs and coronal loops 
appear to be separate entities with magnetic-field structures rooted
in different elements of the same magnetic-flux concentrations.  As
already suggested by several studies, CBPs appear to be composed of
loops at different temperatures \citep[e.g.][]{1990ApJ...352..333H,
  2010ApJ...710.1806D, 2012ApJ...757..167K}.  The interconnection
between coronal and lower-temperature loops and its temporal evolution
has never been explored, possibly because of a lack of suitable
observations.  To identify the precise rooting point of these two
different loop systems requires high-resolution vector magnetic-field
observations and extrapolation models that can successfully cope with both
coronal and chromospheric fields. Such a model is, as mentioned above,
presently under development \citep{2019A&A...631A.162Z}, and it will
be employed in the future when suitable vector magnetograms are
obtained. The HMI vector magnetograms are insufficient owing to the low
signal-to-noise of the transverse component of the field in the quiet
Sun \citep{2014SoPh..289.3483H}.

During the coronal heating episode,   the chromospheric counterpart of the CBP  exhibits a rapid (hydrogen) temperature  increase; specific locations such as a CBP footpoint and a \ha\ loop show a temperature rise of in average $\sim$50\% with respect to the pre-event value.  Also, following the magnetic flux and temperature increase in the corona,  the chromospheric counterpart of the CBP  becomes more dynamic for an interval of about 30~min, which is longer than  the duration of the coronal loop heating.  The typically observed downflows are enhanced possibly by plasma draining from chromospheric, transition-region, and coronal heights. However, upflows, detected as mottles and plasma motions along the CBP loops, also appear to  intensify after the magnetic-flux emergence. Both the hydrogen temperature and the dynamic flows appear to occur with a small delay of less than 3~min after the coronal heating episode has started. This delay together with the identified coronal field structure  suggest the scenario of reconnection-associated coronal heating.   To the best of our knowledge, the only other study that has looked into and reported such a delay is  the paper by \citet{1981SoPh...69...77H} from the Skylab spectroheliogram observations of nine CPBs. The study found a delay of at least 5.5~min (the cadence of their data) between the response in the coronal Mg~{\sc x}~625~\AA, and transition-region O~{\sc vi}~1032~\AA,  C~{\sc ii}~1335~\AA,  and the chromospheric Ly-$\alpha$~1216~\AA\ lines.   

There are several possible sources for the temperature increase in the chromosphere underlying  the CBP loops. The combined effect of the thermal conduction and energetic particle beams would result in a temperature increase in the coronal loop footpoints.  Below temperatures of log T (K) $\ge$ 4.7,  the thermal conduction would be less effective  and heating  by energetic particle beams would prevail. Magnetic reconnection in the corona is known to  generate beams of accelerated particles that are typically observed during solar flares and are known to be responsible for the heating  in the chromospheric footpoints of coronal loops.  However, a very recent study by  \citet{2020A&A...643A..27F} has demonstrated that particle beams can also be a common phenomenon  in the quiet Sun and would largely contribute to the energy transport from the corona into the chromosphere. Another contributor to the footpoint temperature increase  would be shocks caused by the  downflows as typically observed in sunspot plumes (footpoints of large extended loops rooted in sunspot umbras)  for instance \citep[e.g.][]{2014ApJ...789L..42K}.  In addition, the enthalpy flux  due to downflows overwhelms the conductive flux in the low transition region and below, thereby triggering  the loop leg or footpoint temperature increase \citep{1997ApJ...480..817C}.

  We ask how the heating in the coronal loops and their foopoints in the chromosphere could lead to a heating of the chromospheric loops. A  plausible mechanism is through radiative heating, in which  the Ly-$\alpha$ line is the main source  together with heating from EUV optically  thin radiative losses from the transition region and  corona absorbed in the chromosphere \citep[for details see][]{2012A&A...539A..39C,2018A&A...614A.110Z}. To confirm this further numerical simulations are required. A delayed intense increase of the Ly-$\alpha$~1216~\AA\ emission was reported by  \citet{1981SoPh...69...77H} during a heating episode initially recorded in the Mg~{\sc x}~625~\AA\ line.

We should also mention that enhanced nanoflare heating due to magnetic flux increase could  be another possible heating  mechanism. This heating mechanism has already been explored \citep[see section~9 in][]{2019LRSP...16....2M}, but the result remains inconclusive and requires further observational  evidence. Unfortunately, our study cannot provide observational evidence for this mechanism because this would require spectroscopic  observations covering a wide range of formation temperatures combined with nanoflare heating modelling.

\section{Summary and conclusions}
\label{concl}
 
The present study has investigated, for the first time since the work by  \citet{1981SoPh...69...77H}, the temporal variations of  the coronal and chromospheric emission from CBP together with the temporal changes of the magnetic flux in the footpoints of a CBP.  Ground-based \ha\  and \ca\ data are used  to trace changes of both intensities and  Doppler velocities,  and most importantly,  of the hydrogen temperature based on a novel, recently developed inversion technique by \citet{2020A&A...640A..45C}. The observed coronal heating caused by magnetic flux emergence is followed with a small delay by enhanced dynamics and heating of the chromospheric loops and footpoints the CBP. The delay in the response of the chromospheric counterpart of the CBP could suggest that the heating occurs at coronal heights with a consequent heating of the CBP chromospheric counterpart.   An additional heating generated in the  chromosphere, from magnetic reconnection for example, may also be at work but this case study does not provide evidence for this. This scenario will be explored in the future through new CBP multi-temperature observations from ground- and space-based observatories and 3D radiation-MHD   {\it Bifrost}  simulations \citep{2011A&A...531A.154G}. 

\begin{acknowledgements}
We would like to thank very much the anonymous referee for their very helpful suggestions. This work was supported by the National Research Foundation of the Korea (NRF-2019H1D3A2A01099143, NRF-2020R1A2C2004616). FMI is grateful to the Spanish Ministry of Science, Innovation and Universities for support through projects AYA2014-55078-P and PGC2018-095832-B-I00; FMI and DNS gratefully acknowledge the European Research Council (ERC) for the award of the Synergy Grant ``The Whole Sun'' (ERC-2018-SyG 810218). DNS thankfully acknowledges support from the Research Council of Norway through its Centres of Excellence scheme, project number 262622. VAPOR is a product of the National Center for Atmospheric Research's Computational and Information Systems Lab. Support for VAPOR is provided by the U.S. National Science Foundation (grants $\#$ 03-25934 and 09-06379, ACI-14-40412), and by the Korea Institute of Science and Technology Information. The HMI data are provided courtesy of NASA/SDO and corresponding science teams. The HMI data have been retrieved using the Stanford University's Joint Science Operations Centre/Science Data Processing Facility.  We thank Thomas Wiegelmann for providing the LFFF code. The authors thank the ISSI (Bern) for the support to the team ``Observation-Driven Modelling of Solar Phenomena''. 
\end{acknowledgements}

\bibliographystyle{aa}

\begin{thebibliography}{52}
\expandafter\ifx\csname natexlab\endcsname\relax\def\natexlab#1{#1}\fi

\bibitem[{{Andretta} {et~al.}(2012){Andretta}, {Telloni}, \& {Del
  Zanna}}]{2012SoPh..279...53A}
{Andretta}, V., {Telloni}, D., \& {Del Zanna}, G. 2012, \solphys, 279, 53

\bibitem[{{Carlsson} \& {Leenaarts}(2012)}]{2012A&A...539A..39C}
{Carlsson}, M. \& {Leenaarts}, J. 2012, \aap, 539, A39

\bibitem[{{Carlsson} \& {Stein}(1997)}]{1997ApJ...481..500C}
{Carlsson}, M. \& {Stein}, R.~F. 1997, \apj, 481, 500

\bibitem[{{Cauzzi} {et~al.}(2008){Cauzzi}, {Reardon}, {Uitenbroek},
  {Cavallini}, {Falchi}, {Falciani}, {Janssen}, {Rimmele}, {Vecchio}, \&
  {W{\"o}ger}}]{2008A&A...480..515C}
{Cauzzi}, G., {Reardon}, K.~P., {Uitenbroek}, H., {et~al.} 2008, \aap, 480, 515

\bibitem[{{Chae} {et~al.}(2020){Chae}, {Madjarska}, {Kwak}, \&
  {Cho}}]{2020A&A...640A..45C}
{Chae}, J., {Madjarska}, M.~S., {Kwak}, H., \& {Cho}, K. 2020, \aap, 640, A45

\bibitem[{{Chae} {et~al.}(2013){Chae}, {Park}, {Ahn}, {Yang}, {Park}, {Nah},
  {Jang}, {Cho}, {Cao}, \& {Goode}}]{2013SoPh..288....1C}
{Chae}, J., {Park}, H.-M., {Ahn}, K., {et~al.} 2013, \solphys, 288, 1

\bibitem[{{Chae} {et~al.}(2014){Chae}, {Yang}, {Park}, {Ajor Maurya}, {Cho}, \&
  {Yurchysyn}}]{2014ApJ...789..108C}
{Chae}, J., {Yang}, H., {Park}, H., {et~al.} 2014, \apj, 789, 108

\bibitem[{{Chae} {et~al.}(1997){Chae}, {Yun}, \&
  {Poland}}]{1997ApJ...480..817C}
{Chae}, J., {Yun}, H.~S., \& {Poland}, A.~I. 1997, \apj, 480, 817

\bibitem[{{Cheung} \& {Isobe}(2014)}]{2014LRSP...11....3C}
{Cheung}, M. C.~M. \& {Isobe}, H. 2014, Living Reviews in Solar Physics, 11, 3

\bibitem[{{Contarino} {et~al.}(2009){Contarino}, {Zuccarello}, {Romano},
  {Spadaro}, \& {Ermolli}}]{2009A&A...507.1625C}
{Contarino}, L., {Zuccarello}, F., {Romano}, P., {Spadaro}, D., \& {Ermolli},
  I. 2009, \aap, 507, 1625

\bibitem[{{Dacie} {et~al.}(2018){Dacie}, {T{\"o}r{\"o}k}, {D{\'e}moulin},
  {Linton}, {Downs}, {van Driel-Gesztelyi}, {Long}, \&
  {Leake}}]{2018ApJ...862..117D}
{Dacie}, S., {T{\"o}r{\"o}k}, T., {D{\'e}moulin}, P., {et~al.} 2018, \apj, 862,
  117

\bibitem[{{Doschek} {et~al.}(2010){Doschek}, {Landi}, {Warren}, \&
  {Harra}}]{2010ApJ...710.1806D}
{Doschek}, G.~A., {Landi}, E., {Warren}, H.~P., \& {Harra}, L.~K. 2010, \apj,
  710, 1806

\bibitem[{{Frogner} {et~al.}(2020){Frogner}, {Gudiksen}, \&
  {Bakke}}]{2020A&A...643A..27F}
{Frogner}, L., {Gudiksen}, B.~V., \& {Bakke}, H. 2020, \aap, 643, A27

\bibitem[{{Galsgaard} {et~al.}(2017){Galsgaard}, {Madjarska},
  {Moreno-Insertis}, {Huang}, \& {Wiegelmann}}]{2017A&A...606A..46G}
{Galsgaard}, K., {Madjarska}, M.~S., {Moreno-Insertis}, F., {Huang}, Z., \&
  {Wiegelmann}, T. 2017, \aap, 606, A46

\bibitem[{{Galsgaard} {et~al.}(2000){Galsgaard}, {Parnell}, \&
  {Blaizot}}]{Galsgaard_etal_2000}
{Galsgaard}, K., {Parnell}, C.~E., \& {Blaizot}, J. 2000, \aap, 362, 395

\bibitem[{{Golding} {et~al.}(2017){Golding}, {Leenaarts}, \&
  {Carlsson}}]{2017A&A...597A.102G}
{Golding}, T.~P., {Leenaarts}, J., \& {Carlsson}, M. 2017, \aap, 597, A102

\bibitem[{{Gudiksen} {et~al.}(2011){Gudiksen}, {Carlsson}, {Hansteen}, {Hayek},
  {Leenaarts}, \& {Mart{\'\i}nez-Sykora}}]{2011A&A...531A.154G}
{Gudiksen}, B.~V., {Carlsson}, M., {Hansteen}, V.~H., {et~al.} 2011, \aap, 531,
  A154

\bibitem[{{Habbal} \& {Withbroe}(1981)}]{1981SoPh...69...77H}
{Habbal}, S.~R. \& {Withbroe}, G.~L. 1981, \solphys, 69, 77

\bibitem[{{Habbal} {et~al.}(1990){Habbal}, {Withbroe}, \&
  {Dowdy}}]{1990ApJ...352..333H}
{Habbal}, S.~R., {Withbroe}, G.~L., \& {Dowdy}, Jr., J.~F. 1990, \apj, 352, 333

\bibitem[{{Hansteen} {et~al.}(2019){Hansteen}, {Ortiz}, {Archontis},
  {Carlsson}, {Pereira}, \& {Bj{\o}rgen}}]{2019A&A...626A..33H}
{Hansteen}, V., {Ortiz}, A., {Archontis}, V., {et~al.} 2019, \aap, 626, A33

\bibitem[{{Haynes} \& {Parnell}(2007)}]{2007PhPl...14h2107H}
{Haynes}, A.~L. \& {Parnell}, C.~E. 2007, Physics of Plasmas, 14, 082107

\bibitem[{{Hoeksema} {et~al.}(2014){Hoeksema}, {Liu}, {Hayashi}, {Sun},
  {Schou}, {Couvidat}, {Norton}, {Bobra}, {Centeno}, {Leka}, {Barnes}, \&
  {Turmon}}]{2014SoPh..289.3483H}
{Hoeksema}, J.~T., {Liu}, Y., {Hayashi}, K., {et~al.} 2014, \solphys, 289, 3483

\bibitem[{{Kayshap} \& {Dwivedi}(2017)}]{2017SoPh..292..108K}
{Kayshap}, P. \& {Dwivedi}, B.~N. 2017, \solphys, 292, 108

\bibitem[{{Kleint} {et~al.}(2014){Kleint}, {Antolin}, {Tian}, {Judge}, {Testa},
  {De Pontieu}, {Mart{\'\i}nez-Sykora}, {Reeves}, {Wuelser}, {McKillop},
  {Saar}, {Carlsson}, {Boerner}, {Hurlburt}, {Lemen}, {Tarbell}, {Title},
  {Golub}, {Hansteen}, {Jaeggli}, \& {Kankelborg}}]{2014ApJ...789L..42K}
{Kleint}, L., {Antolin}, P., {Tian}, H., {et~al.} 2014, \apjl, 789, L42

\bibitem[{{Kwon} {et~al.}(2012){Kwon}, {Chae}, {Davila}, {Zhang}, {Moon},
  {Poomvises}, \& {Jones}}]{2012ApJ...757..167K}
{Kwon}, R.-Y., {Chae}, J., {Davila}, J.~M., {et~al.} 2012, \apj, 757, 167

\bibitem[{{Leenaarts} {et~al.}(2012){Leenaarts}, {Carlsson}, \& {Rouppe van der
  Voort}}]{2012ApJ...749..136L}
{Leenaarts}, J., {Carlsson}, M., \& {Rouppe van der Voort}, L. 2012, \apj, 749,
  136

\bibitem[{{Lemen} {et~al.}(2012){Lemen}, {Title}, {Akin}, {Boerner}, {Chou},
  {Drake}, {Duncan}, {Edwards}, {Friedlaender}, {Heyman}, {Hurlburt}, {Katz},
  {Kushner}, {Levay}, {Lindgren}, {Mathur}, {McFeaters}, {Mitchell}, {Rehse},
  {Schrijver}, {Springer}, {Stern}, {Tarbell}, {Wuelser}, {Wolfson}, {Yanari},
  {Bookbinder}, {Cheimets}, {Caldwell}, {Deluca}, {Gates}, {Golub}, {Park},
  {Podgorski}, {Bush}, {Scherrer}, {Gummin}, {Smith}, {Auker}, {Jerram},
  {Pool}, {Soufli}, {Windt}, {Beardsley}, {Clapp}, {Lang}, \&
  {Waltham}}]{2012SoPh..275...17L}
{Lemen}, J.~R., {Title}, A.~M., {Akin}, D.~J., {et~al.} 2012, \solphys, 275, 17

\bibitem[{{Longcope}(1998)}]{Longcope_1998}
{Longcope}, D.~W. 1998, \apj, 507, 433

\bibitem[{{Madjarska}(2019)}]{2019LRSP...16....2M}
{Madjarska}, M.~S. 2019, Living Reviews in Solar Physics, 16, 2

\bibitem[{{Madjarska} {et~al.}(2003){Madjarska}, {Doyle}, {Teriaca}, \&
  {Banerjee}}]{2003A&A...398..775M}
{Madjarska}, M.~S., {Doyle}, J.~G., {Teriaca}, L., \& {Banerjee}, D. 2003,
  \aap, 398, 775

\bibitem[{{Madjarska} {et~al.}(2020){Madjarska}, {Galsgaard}, {Mackay},
  {Koleva}, \& {Dechev}}]{2020A&A...643A..19M}
{Madjarska}, M.~S., {Galsgaard}, K., {Mackay}, D.~H., {Koleva}, K., \&
  {Dechev}, M. 2020, \aap, 643, A19

\bibitem[{{Morgan} \& {Druckm{\"u}ller}(2014)}]{2014SoPh..289.2945M}
{Morgan}, H. \& {Druckm{\"u}ller}, M. 2014, \solphys, 289, 2945

\bibitem[{{Mou} {et~al.}(2018){Mou}, {Madjarska}, {Galsgaard}, \&
  {Xia}}]{2018A&A...619A..55M}
{Mou}, C., {Madjarska}, M.~S., {Galsgaard}, K., \& {Xia}, L. 2018, \aap, 619,
  A55

\bibitem[{{N{\'o}brega-Siverio} {et~al.}(2017){N{\'o}brega-Siverio},
  {Mart{\'\i}nez-Sykora}, {Moreno-Insertis}, \& {Rouppe van der
  Voort}}]{2017ApJ...850..153N}
{N{\'o}brega-Siverio}, D., {Mart{\'\i}nez-Sykora}, J., {Moreno-Insertis}, F.,
  \& {Rouppe van der Voort}, L. 2017, \apj, 850, 153

\bibitem[{{O'Dwyer} {et~al.}(2010){O'Dwyer}, {Del Zanna}, {Mason}, {Weber}, \&
  {Tripathi}}]{2010A&A...521A..21O}
{O'Dwyer}, B., {Del Zanna}, G., {Mason}, H.~E., {Weber}, M.~A., \& {Tripathi},
  D. 2010, \aap, 521, A21

\bibitem[{{Parnell} {et~al.}(1994){Parnell}, {Priest}, \&
  {Golub}}]{Parnell_etal_1994_3d}
{Parnell}, C.~E., {Priest}, E.~R., \& {Golub}, L. 1994, \solphys, 151, 57

\bibitem[{{Pesnell} {et~al.}(2012){Pesnell}, {Thompson}, \&
  {Chamberlin}}]{2012SoPh..275....3P}
{Pesnell}, W.~D., {Thompson}, B.~J., \& {Chamberlin}, P.~C. 2012, \solphys,
  275, 3

\bibitem[{{Priest} {et~al.}(1994){Priest}, {Parnell}, \&
  {Martin}}]{Priest_etal_1994}
{Priest}, E.~R., {Parnell}, C.~E., \& {Martin}, S.~F. 1994, \apj, 427, 459

\bibitem[{{Rutten}(1999)}]{1999ASPC..184..181R}
{Rutten}, R.~J. 1999, Astronomical Society of the Pacific Conference Series,
  Vol. 184, {(Inter-),Network Structure and DynamicS}, ed. B.~{Schmieder},
  A.~{Hofmann}, \& J.~{Staude}, 181--200

\bibitem[{{Rutten} \& {Uitenbroek}(1991)}]{1991SoPh..134...15R}
{Rutten}, R.~J. \& {Uitenbroek}, H. 1991, \solphys, 134, 15

\bibitem[{{Scherrer} {et~al.}(2012){Scherrer}, {Schou}, {Bush}, {Kosovichev},
  {Bogart}, {Hoeksema}, {Liu}, {Duvall}, {Zhao}, {Title}, {Schrijver},
  {Tarbell}, \& {Tomczyk}}]{2012SoPh..275..207S}
{Scherrer}, P.~H., {Schou}, J., {Bush}, R.~I., {et~al.} 2012, \solphys, 275,
  207

\bibitem[{{Tian} {et~al.}(2008){Tian}, {Curdt}, {Marsch}, \&
  {He}}]{2008ApJ...681L.121T}
{Tian}, H., {Curdt}, W., {Marsch}, E., \& {He}, J. 2008, \apjl, 681, L121

\bibitem[{{Tsiropoula} {et~al.}(2012){Tsiropoula}, {Tziotziou}, {Kontogiannis},
  {Madjarska}, {Doyle}, \& {Suematsu}}]{2012SSRv..169..181T}
{Tsiropoula}, G., {Tziotziou}, K., {Kontogiannis}, I., {et~al.} 2012, \ssr,
  169, 181

\bibitem[{{von Rekowski} {et~al.}(2006{\natexlab{a}}){von Rekowski}, {Parnell},
  \& {Priest}}]{Von_rekowski_etal_1_2006}
{von Rekowski}, B., {Parnell}, C.~E., \& {Priest}, E.~R. 2006{\natexlab{a}},
  \mnras, 366, 125

\bibitem[{{von Rekowski} {et~al.}(2006{\natexlab{b}}){von Rekowski}, {Parnell},
  \& {Priest}}]{Von_rekowski_etal_2_2006}
{von Rekowski}, B., {Parnell}, C.~E., \& {Priest}, E.~R. 2006{\natexlab{b}},
  \mnras, 369, 43

\bibitem[{{Wiegelmann} \& {Neukirch}(2002)}]{2002SoPh..208..233W}
{Wiegelmann}, T. \& {Neukirch}, T. 2002, \solphys, 208, 233

\bibitem[{{Wiegelmann} {et~al.}(2010){Wiegelmann}, {Solanki}, {Borrero},
  {Mart{\'\i}nez Pillet}, {del Toro Iniesta}, {Domingo}, {Bonet}, {Barthol},
  {Gandorfer}, {Kn{\"o}lker}, {Schmidt}, \& {Title}}]{2010ApJ...723L.185W}
{Wiegelmann}, T., {Solanki}, S.~K., {Borrero}, J.~M., {et~al.} 2010, \apjl,
  723, L185

\bibitem[{{Wyper} {et~al.}(2018){Wyper}, {DeVore}, {Karpen}, {Antiochos}, \&
  {Yeates}}]{2018ApJ...864..165W}
{Wyper}, P.~F., {DeVore}, C.~R., {Karpen}, J.~T., {Antiochos}, S.~K., \&
  {Yeates}, A.~R. 2018, \apj, 864, 165

\bibitem[{{Xia} {et~al.}(2003){Xia}, {Marsch}, \&
  {Curdt}}]{2003A&A...399L...5X}
{Xia}, L.~D., {Marsch}, E., \& {Curdt}, W. 2003, \aap, 399, L5

\bibitem[{{Xie} {et~al.}(2017){Xie}, {Madjarska}, {Li}, {Huang}, {Xia},
  {Wiegelmann}, {Fu}, \& {Mou}}]{2017ApJ...842...38X}
{Xie}, H., {Madjarska}, M.~S., {Li}, B., {et~al.} 2017, \apj, 842, 38

\bibitem[{{Zacharias} {et~al.}(2018){Zacharias}, {Hansteen}, {Leenaarts},
  {Carlsson}, \& {Gudiksen}}]{2018A&A...614A.110Z}
{Zacharias}, P., {Hansteen}, V.~H., {Leenaarts}, J., {Carlsson}, M., \&
  {Gudiksen}, B.~V. 2018, \aap, 614, A110

\bibitem[{{Zhu} \& {Wiegelmann}(2019)}]{2019A&A...631A.162Z}
{Zhu}, X. \& {Wiegelmann}, T. 2019, \aap, 631, A162

\end{thebibliography}

\begin{appendix}
\section{On-line material}

\begin{figure}[!ht]
\ifnum\inclfigs>0
\hbox{
\includegraphics[scale=0.15]{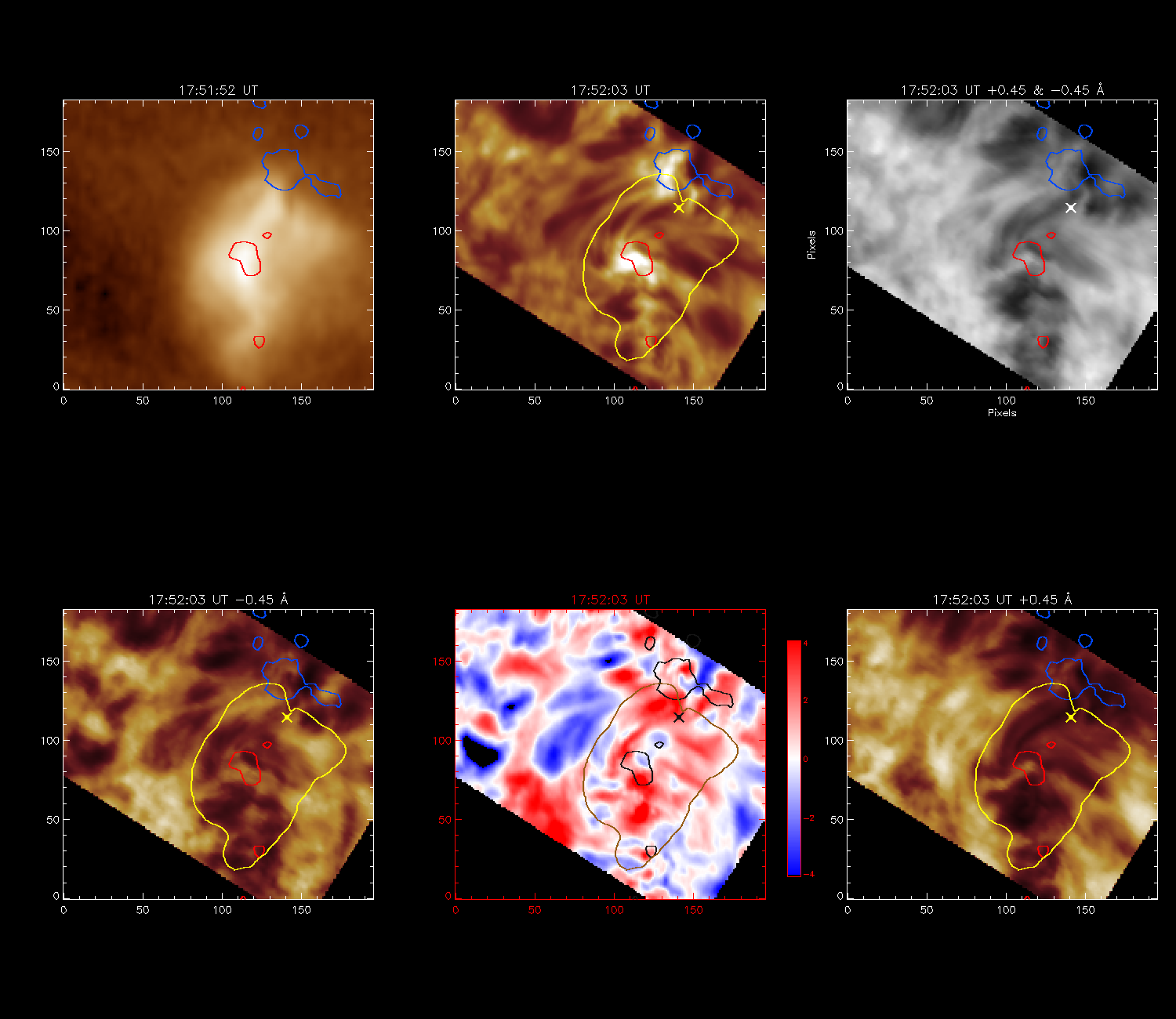}}
\caption{Animation of the following image sequences. From left to right in the top row:   \aa\ image, \ha\ line-core image, and \ha\ image produced from the sum of the intensity in the blue and red wing of \ha\ at $\pm$45~\AA. Bottom row: \ha\ image at $-$0.45~\AA, Doppler-shift image, and \ha\ image at 0.45~\AA. The contour is the CBP \aa\ intensity contour. The $\pm$50~G magnetic-field contours (red -- positive polarity; blue -- negative). See movie\_ha.mov.}
\label{figa1}
\end{figure}

\begin{figure}[!ht]
\ifnum\inclfigs>0
\includegraphics[scale=0.15]{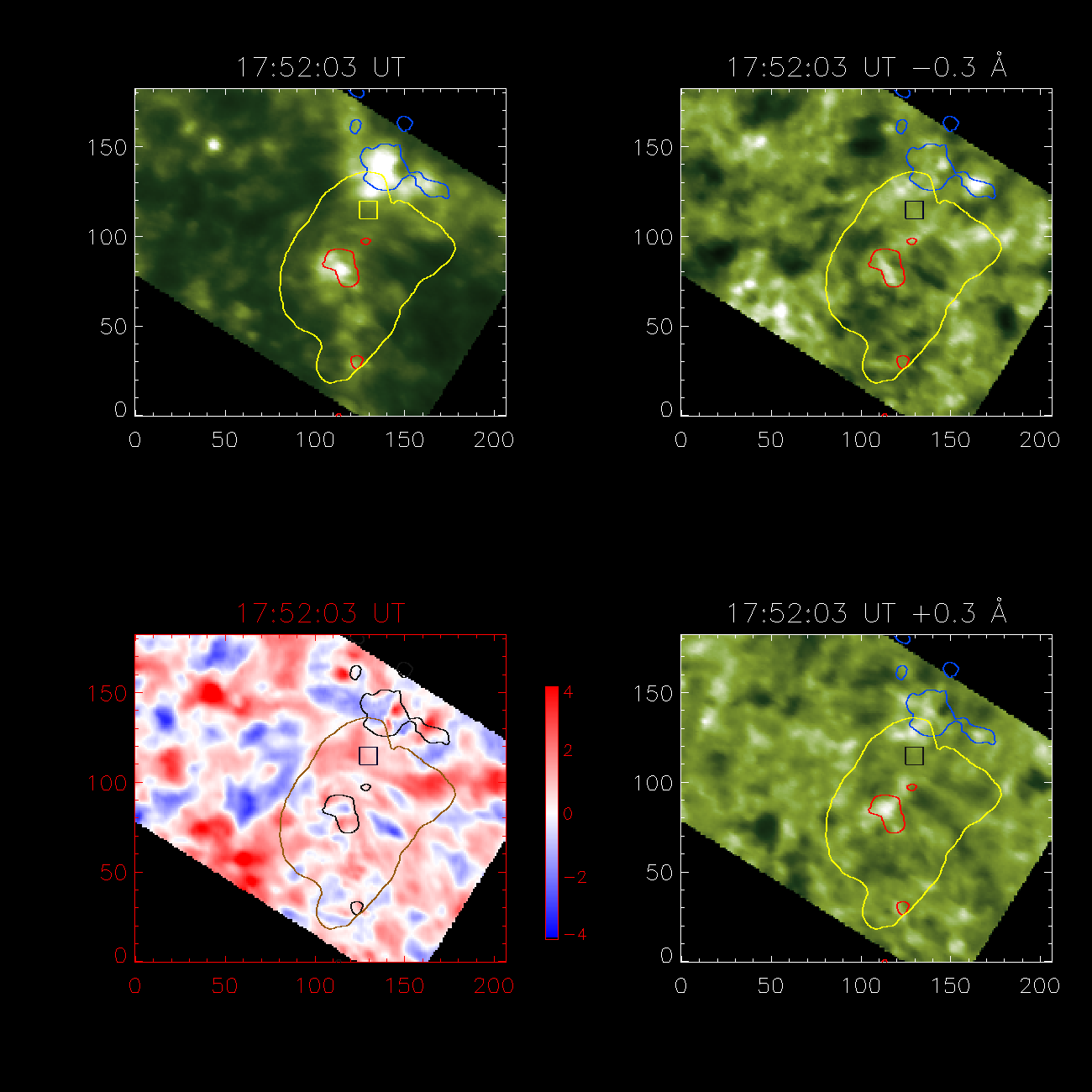}
\fi
\caption{Animation of the following image sequences. Top row: \ca\  line-core  intensity image (left) and \ca\ image at $-$0.3~\AA\ (right). Bottom row: Doppler-shift image and  \ca\ image at 0.3~\AA. The contour overplotted on all panels is the CBP \aa\ intensity contour. The $\pm$50~G magnetic-field contours (red -- positive polarity; blue -- negative). See movie\_ca.mov.}
\label{figa2}
\end{figure}

\begin{figure}[!ht]
\ifnum\inclfigs>0
\includegraphics[scale=0.15]{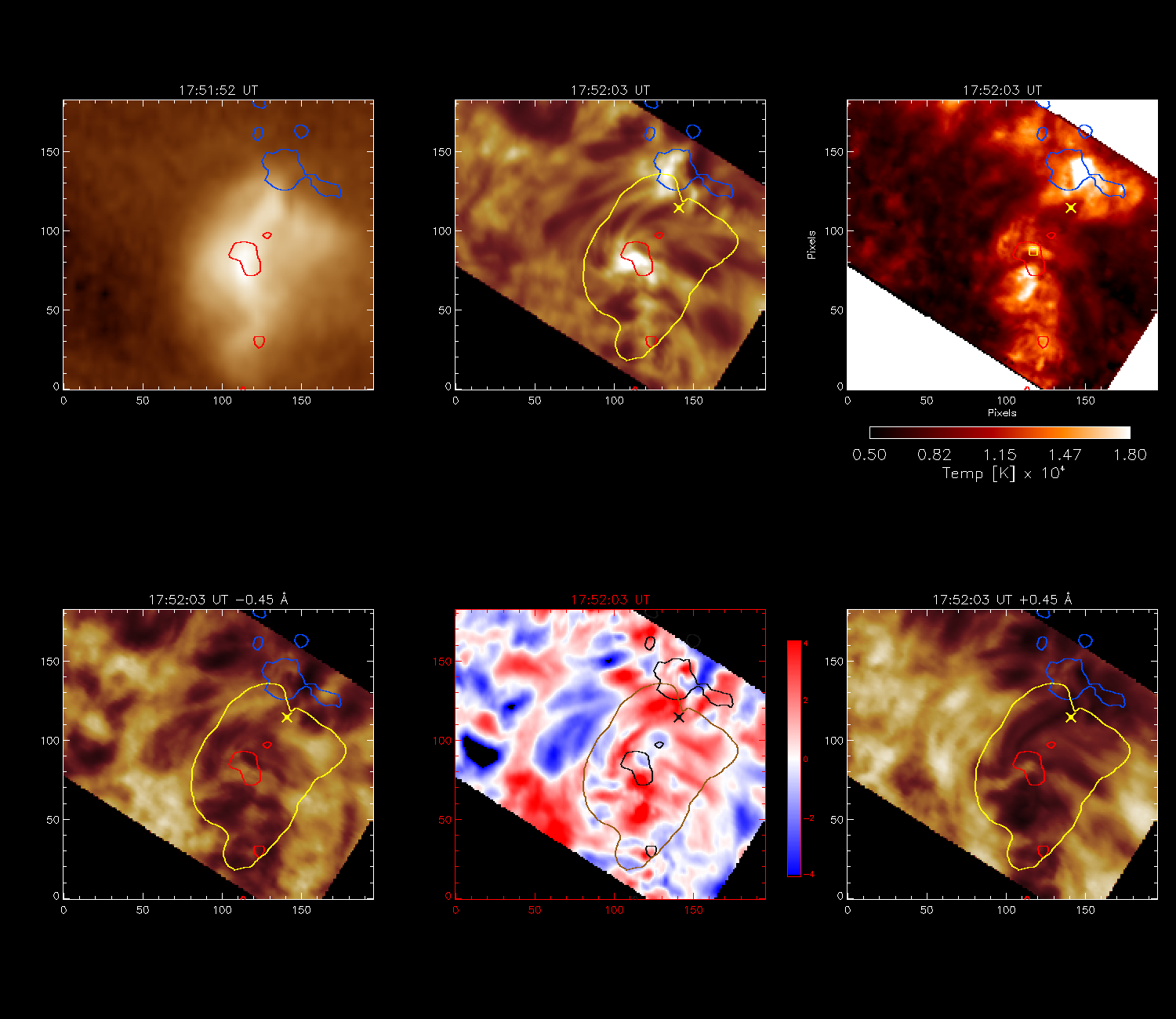}
\fi
\caption{Animation of the following image sequences. From left to right in the top row:  \aa\ image, \ha\ line-core image, and hydrogen temperature image. Bottom row: \ha\ image at $-$0.45~\AA, Doppler-shift image, and \ha\ image at +0.45~\AA. The contour  is the CBP \aa\ intensity contour. The $\pm$50~G magnetic-field contours (red -- positive polarity; blue -- negative). See movie\_temp.mov.}
\label{figa3}
\end{figure}

\begin{figure}[!ht]
\ifnum\inclfigs>0
\includegraphics[scale=0.60]{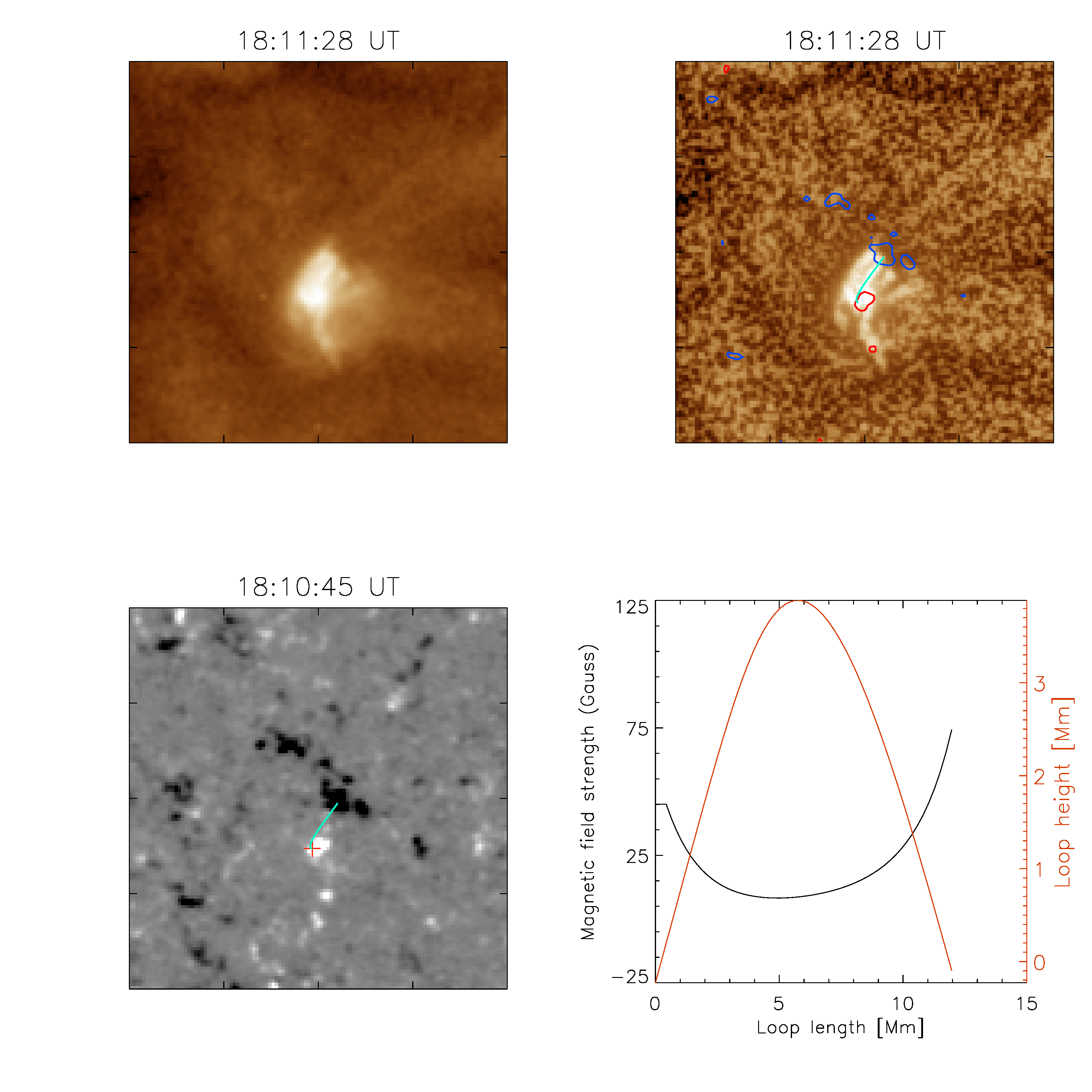}
\fi
\caption{Linear force-free field extrapolation of a CBP loop. Top left: \aa\ image. Top right: An MGN processed \aa\ image overplotted  with a  magnetic field line that matches one of the extrapolated CBP loops. Bottom left: An HMI magnetogram overplotted with the same magnetic field line. The cross indicates the location from which the magnetic-field line was traced. Bottom right: Magnetic-field-line  parameters.}
\label{figa4}
\end{figure}

\end{appendix}

\end{document}